\documentclass{article} 
\usepackage{amsmath,latexsym,graphicx,psfrag,graphicx,caption2}
\usepackage{a4}
\usepackage{cite}  
\usepackage{epsfig}  
\def\d{\hbox{d}} 
\def\lapprox{\lower .7ex\hbox{$\;\stackrel{\textstyle <}{\sim}\;$}}
\def\gapprox{\lower .7ex\hbox{$\;\stackrel{\textstyle >}{\sim}\;$}}

\renewcommand {\d}{\mathrm{d}}
\newcommand {\G}{\tilde g}
\newcommand {\Q}{\tilde q}

\begin{document} 
\unitlength1cm 
\begin{titlepage} 
\vspace*{-1cm} 
\begin{flushright} 
ZU--TH 10/04\\
hep-ph/0406222\\
June 2004
\end{flushright} 
\vskip 3.5cm 

\begin{center} 
{\Large\bf Spin Asymmetries in Squark and Gluino Production\\[2mm] at 
Polarized Hadron Colliders}
\vskip 1.cm 
{\large  T.~Gehrmann, D.~Ma\^{\i}tre}
and {\large D.~Wyler}
\vskip .7cm 
{\it Institut f\"ur Theoretische Physik, Universit\"at Z\"urich,
Winterthurerstrasse 190,\\ CH-8057 Z\"urich, Switzerland} 
\end{center} 
\vskip 2.6cm 

\begin{abstract} 
We study the production cross sections for squarks and gluinos 
in collision of longitudinally polarized hadrons. The corresponding 
polarized partonic cross sections are computed in leading order supersymmetric 
QCD. The resulting asymmetries are evaluated for 
the polarized proton collider RHIC, as well as for hypothetical polarized 
options of the Tevatron and the LHC. These asymmetries turn out to be 
sizable over a wide range of supersymmetric particle masses. 
Once supersymmetric particles are discovered in unpolarized 
collisions, a measurement of the spin asymmetries would thus 
potentially help to establish the properties of the newly discovered particles
and open a window  to detailed sparticle spectroscopy at future polarized 
hadron colliders. 
\end{abstract} 
\vfill 
\end{titlepage} 
\newpage

\section{Introduction}
Data from inclusive and semi-inclusive 
deep inelastic scattering off polarized 
targets~\cite{smc,slac,hermes,compass,jha} and from
upcoming  polarized proton-proton
collisions at RHIC~\cite{rhic} continue to improve on our knowledge on
the polarized parton distributions in the nucleon. They
parametrise the probability of finding 
partons inside the nucleon having 
their spin aligned or anti-aligned to the nucleon spin and are obtained from 
a global fit~\cite{grsvo,gs,abfr,grsv,aac,bb}
to spin asymmetries in polarized lepton-hadron and hadron-hadron 
collisions. They can be used 
to predict cross sections in collisions of hadrons with fixed 
spin orientation. 

In the near future, one could therefore envisage to study 
polarized hadronic collisions not only to probe 
the structure of the colliding hadrons, but also to 
employ the polarization information in searches for new physics
beyond the standard model. So far, there is very little theoretical work 
in this direction. The only comprehensive studies 
of new physics 
searches in polarized hadronic collisions to date concern
the effects of contact interactions~\cite{virey,virey3}
and of new gauge bosons~\cite{virey2} on spin asymmetries at 
RHIC and HERA. It was found that polarization information 
can yield vital insights, such as discriminating different types of 
contact interactions that yield identical signatures at unpolarized colliders. 
Likewise, one would expect that the additional information 
would strongly improve the study of other new physics scenarios.

Probably the most promising scenarios for physics 
beyond the standard model is its supersymmetric extension, formulated 
in the minimal supersymmetric standard model (MSSM)~\cite{mssm}. 
The effects of beam polarization in supersymmetric particle production 
have been studied in detail~\cite{gudi,moretti} 
for a future linear electron--positron 
collider~\cite{tesla}.  
In the same context, fermion pair production in polarized photon-photon 
collisions~\cite{klasen1} and
gluino pair production in polarized electron-position 
and polarized direct photon-photon  
collisions have been examined~\cite{klasen}.
In all these studies, it was demonstrated that beam polarization asymmetries 
could be helpful in determining several parameters of the 
MSSM~\cite{gudi2} .

Spin asymmetries in supersymmetric particle production at hadron colliders
have so far been considered in early studies adressing particular partonic
production channels for squark and gluino production~\cite{ratcliffe} and
scalar lepton production~\cite{soffer}.

On the other hand, the production of squarks and gluinos at hadron colliders is
a QCD process and depends only
%interactions of the squark and gluino colour 
%charges. These cross sections are therefore independent the parameters of 
%the MSSM, 
on the squark and gluino masses, thus allowing a 
precise prediction~\cite{susylo} independent of other parameters of the MSSM.
The expected cross sections 
for present and future hadron colliders turn out to be sizable and 
are further enhanced by QCD next-to-leading order corrections~\cite{been}. 
Non-observation of squark and gluino signatures at the Tevatron thus 
turns into stringent limits on the squark and gluino masses in the framework 
of the MSSM~\cite{tevdata}: $m_{\tilde{q}} > 250$~GeV, 
$m_{\tilde{g}}> 195$~GeV.

These limits are substantially 
weakened if more complicated supersymmetric models than the MSSM are 
considered. In such non-minimal realizations of supersymmetry, squark and 
gluino production is also mediated by QCD interactions, thus yielding the 
same production cross sections as in the MSSM. Unfortunately, the non-standard 
decay modes of squarks and gluinos in these models cannot be distinguished
from background processes at unpolarized colliders. 
One such model was for example advocated to explain an apparent 
excess in the  bottom production cross section at the 
Tevatron~\cite{berger}, predicting 
gluino and squark masses  as low as several GeV.
\footnote{Data from electron-positron annihilation~\cite{janot},
improved understanding of bottom quark 
production in the standard model~\cite{cacciari} and new Tevatron
data~\cite{newb} not exposing the excess all make this model already
unlikely now, although there is still discussion in the literature 
\cite{nadolsky} that this mass range can fully ruled out.}

In the present paper, we  calculate the  spin asymmetries 
in squark and gluino production at polarized hadron colliders.
At present, there is only RHIC at BNL where 
two polarized proton beams with a maximum centre-of-mass energy 
of $\sqrt{s} = 500$~GeV collide. Obviously, this is not sufficient to produce 
the MSSM sparticles; however they could be within reach 
if supersymmetry is realized in a more exotic scenario. 
To illustrate the potential of a future polarized hadron collider with a
centre-of-mass energy comparable to Tevatron or LHC, we also 
study the anticipated spin asymmetries at polarized versions of Tevatron and 
LHC. Up to now, beam polarization has not been considered as a 
realistic upgrade option for these colliders, in part because the 
physics potential of such an option was not studied. With this paper we 
also wish to direct attention to this possibility.

The paper is organised as follows. In Section~\ref{sec:cross}, we derive 
the spin-dependent parton level cross sections for squark and gluino 
production at hadron colliders in leading order QCD. Section~\ref{sec:num}
contains numerical results for the expected cross sections, 
spin asymmetries and the statistical accuracy with which they could 
be measured 
at RHIC and at polarized versions of Tevatron and LHC. 
To gain more detailed insight into the spin asymmetries, we discuss their 
dependence on the transverse momentum and the rapidity of the squarks and 
gluinos in Section~\ref{sec:pt}. Finally, some conclusions and an outlook
are presented in Section~\ref{sec:conc}. 

\section{Polarized cross sections}
\label{sec:cross}
  The leading order (LO) processes involved in squark and gluino production in
  a proton-proton (LHC, RHIC) or a proton-antiproton collider (Tevatron) are 
\begin{itemize}
  \item[-] for squark-antisquark pair production 
  \begin{eqnarray}  \nonumber
  q+\bar q &\rightarrow &\tilde q+ \bar{\tilde q}\nonumber  \\ g+g
  &\rightarrow&	\tilde q +\bar{\tilde q}\nonumber, 
   \end{eqnarray}
  \item[-] for squark-squark (antisquark-antisquark) production
  \begin{eqnarray}  \nonumber
  q+q &\rightarrow& \tilde q+ \tilde q \\ \bar q+\bar q &\rightarrow&
  \bar{\tilde q}+\bar{\tilde q},\nonumber 
   \end{eqnarray}
   \item[-] for gluino-gluino production
\begin{eqnarray}  \nonumber
  g+g &\rightarrow &\tilde g+ \tilde g \\ q+ \bar q &\rightarrow&	\tilde
  g +\tilde g\nonumber, 
   \end{eqnarray}
\item[-] for squark-gluino (antisquark-gluino) production
\begin{eqnarray}  \nonumber
  q+g &\rightarrow& \tilde q+\tilde g \nonumber\\ \bar q +g
  &\rightarrow&\bar{\tilde q} +\tilde g \nonumber. 
\end{eqnarray}
\end{itemize}
The charge conjugated processes (in parentheses) will not be written out 
explicitly, they are included in the numerical results where appropriate. 
The matrix elements are calculated using the Feynman rules of SUSY QCD 
listed in~\cite{rosiek}. The fermion number violating interactions induced 
by the gluinos are
treated according to ref. \cite{ferm_viol}. 

Incoming quarks (including incoming $b$ quarks) are assumed to be massless,
such that we have $n_f=5$ light flavours. We only consider final state 
squarks corresponding to the light quark flavours. All 
squark masses are taken equal to $m_{\tilde q}$ 
\footnote{$L$-squarks and $R$-squarks are therefore mass-degenerate  
and experimentally indistinguishable.}. We do not consider in detail top squark
production where these assumptions do not hold and which require
a more dedicated treatment~\cite{plehn}.

To describe the process kinematics, we assign momenta $k_1$ and 
$k_2$ to the incoming partons and $p_1$ and $p_2$ to 
the outgoing squarks and gluinos. We introduce the Mandelstam variables 
\begin{eqnarray}
s&=&(k_1+k_2)^2=+2k_1\cdot k_2 \nonumber \\
t&=&(k_2-p_2)^2=-2k_2 \!\cdot\! p_2 +p_2^2\nonumber \\
u&=&(k_1-p_2)^2=-2k_1 \!\cdot\! p_2 +p_2^2\nonumber \;,
\end{eqnarray}
which satisfy $s+t+u=p_1^2+p_2^2$.
The following variables are also useful~\cite{been}:
\begin{eqnarray}
t_1&=&(k_2-p_2)^2-m_{\tilde q}^2 \;,\qquad 
t_g=(k_2-p_2)^2-m_{\tilde g}^2\;, \nonumber\\
u_1&=&(k_1-p_2)^2-m_{\tilde q}^2 \;,\qquad 
u_g=(k_1-p_2)^2-m_{\tilde g}^2 \nonumber.
\end{eqnarray}
The colour factors corresponding to different production processes are 
$C_F=(N^2-1)/2N$, $C_O=N(N^2-1)$ and $C_K=(N^2-1)/N$, 
with $N=3$. 
%The strong coupling constant is $\alpha_s = g_s^2/(4\pi)$. 

To extract the dependence on the helicities $\lambda$, 
we introduce helicity projections
for spinors and gluons. 
%Since we are interested in the dependence of the production cross section 
%on the helicities of the incoming partons, we have to introduce 
%helicity projectors in the evaluation of the squared matrix elements.
For the spinors we take the usual form
%the  the projection on the helicity $\lambda$ state is
%is  achieved by multiplying 
%the incoming quark spinor with the helicity projection operator
\begin{equation}
\frac{1}{2} \left(1 + \lambda \gamma_5  \right)\; ,
\end{equation}
but more care is needed for the incoming gluons. We use an 
axial gauge to ensure that only physical polarization states 
contribute to the cross section. In this gauge, the product of the polarization
vector of an incoming parton with momentum
$k$ and its complex conjugate can be written as
\begin{equation}
\epsilon_\mu^\pm(k,q)\cdot(\epsilon_\nu^\pm(k,q))^*
=\epsilon_\mu^\pm(k,q)\cdot\epsilon_\nu^\mp(k,q)=\frac{1}{2k\cdot q}
{\rm Tr}\left(\frac{1\mp\gamma^5}{2}\gamma_\mu\not k\gamma_\nu\not q\right),  
\end{equation}
where the 'gauge vector' $q$ is chosen conveniently. This form is
independent of the representation of $\gamma_5$. 

Corresponding to the helicities $\lambda_{1,2}$ of
the incoming partons, there are four different squared matrix elements
\begin{equation}
|{\cal M}|^2_{\lambda_1,\lambda_2} ;\,
\end{equation}
where $\lambda_{1,2} =-1, 1$ correspond partons with negative and positive
helicity. 
If the final state squark chiralities are summed over, parity invariance of 
SUSY QCD yields 
\begin{equation}
|{\cal M}|^2_{1,-1} =|{\cal M}|^2_{-1,1} \qquad \mbox{and} \qquad
|{\cal M}|^2_{-1,-1} =|{\cal M}|^2_{1,1}\;.
\end{equation}
Therefore the full polarization dependence is given by the 
sum and the difference
\begin{eqnarray}
|{\cal M}|^2 &=&  |{\cal M}|^2_{1,1} + |{\cal M}|^2_{1,-1} + 
|{\cal M}|^2_{-1,1} + |{\cal M}|^2_{-1,-1} \; ,\\
\Delta |{\cal M}|^2 &=&  |{\cal M}|^2_{1,1} - |{\cal M}|^2_{1,-1} 
-|{\cal M}|^2_{-1,1} + |{\cal M}|^2_{-1,-1} \;,
\end{eqnarray}
which we compute at leading order for all partonic squark and 
gluino production processes.

\subsection{Polarized squared matrix elements}
\begin{figure}[tb]
\begin{center}
\epsfig{file=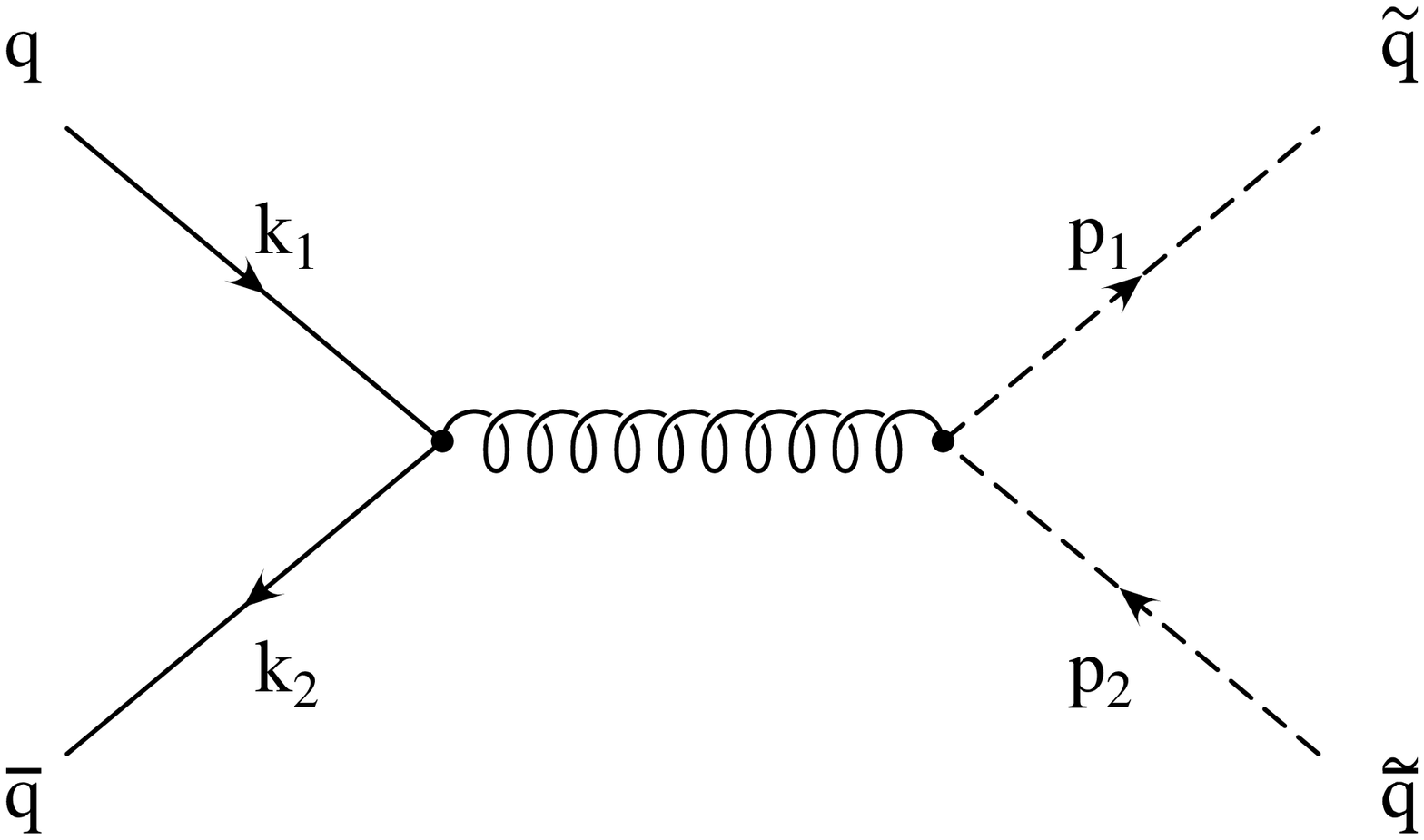,width=4cm}
\epsfig{file=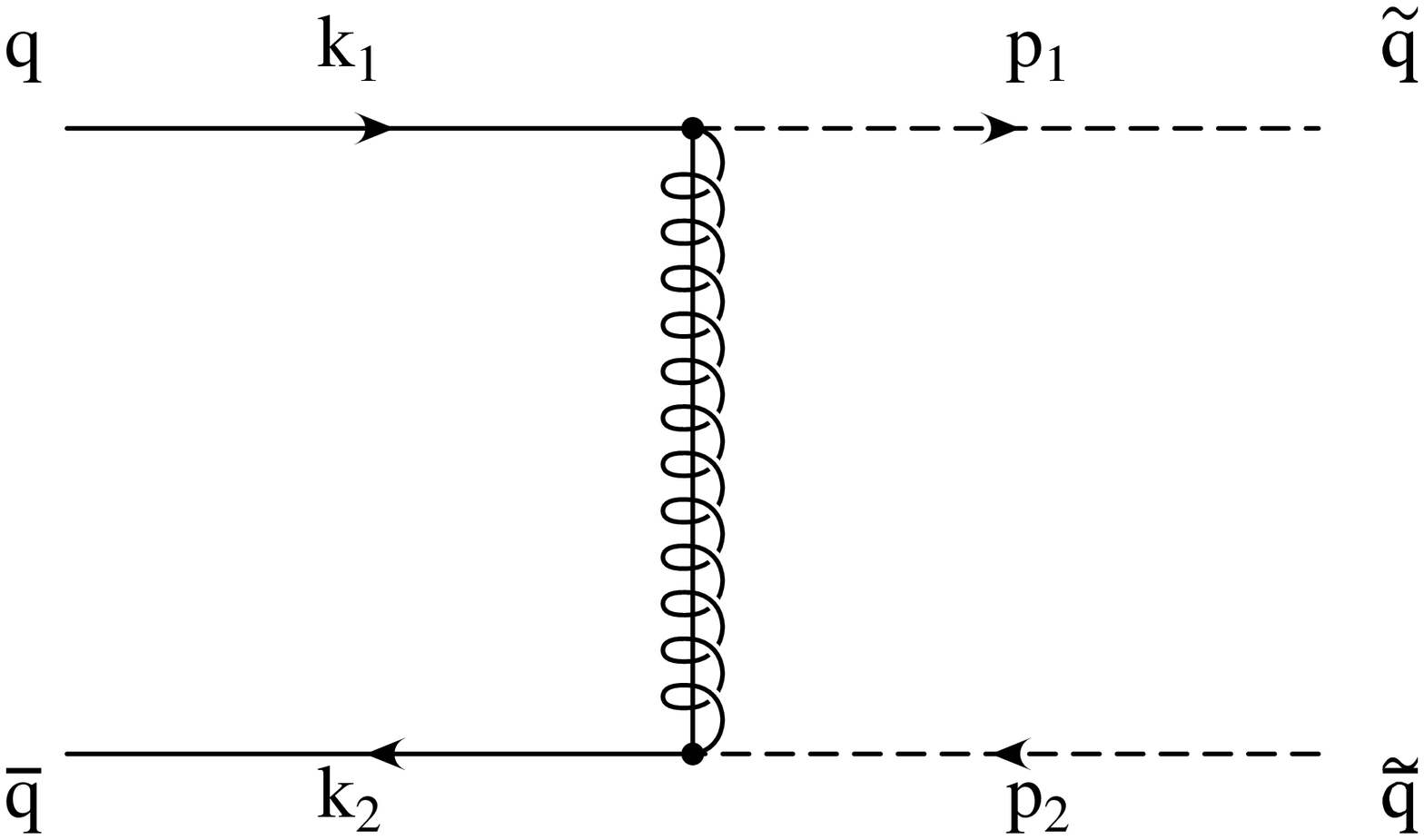,width=4cm}\\ (a) \\
\epsfig{file=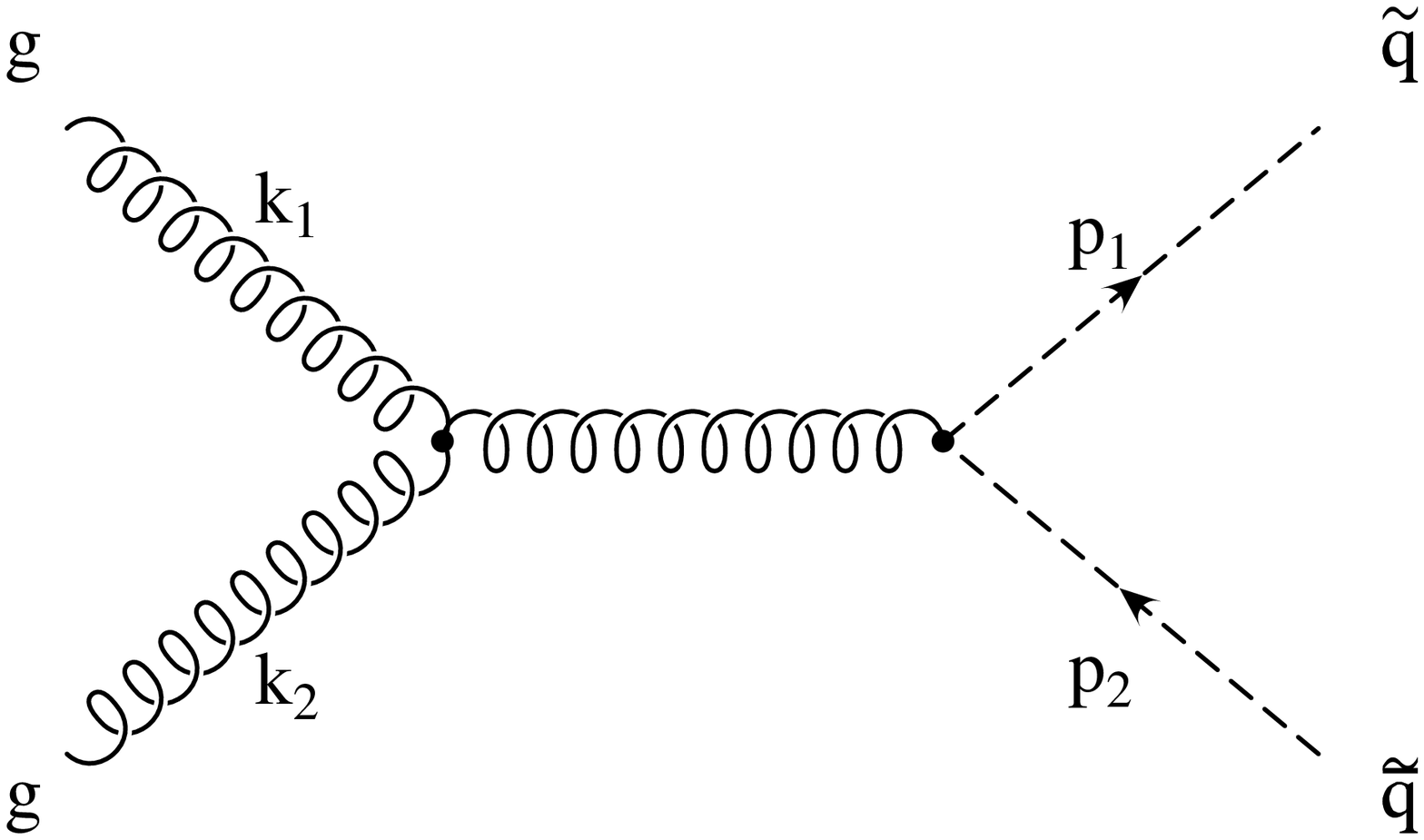,width=4cm}
\epsfig{file=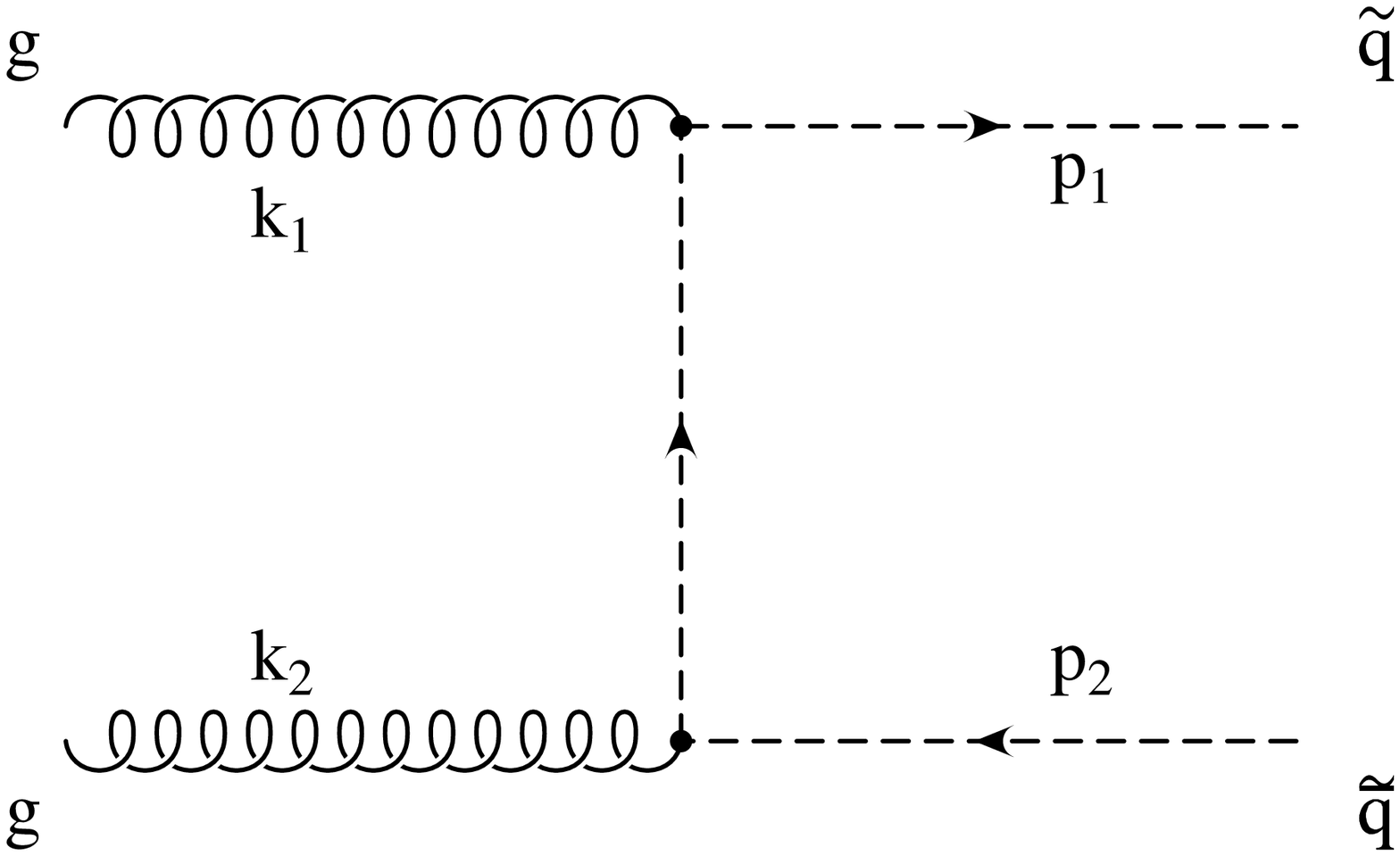,width=4cm}
\epsfig{file=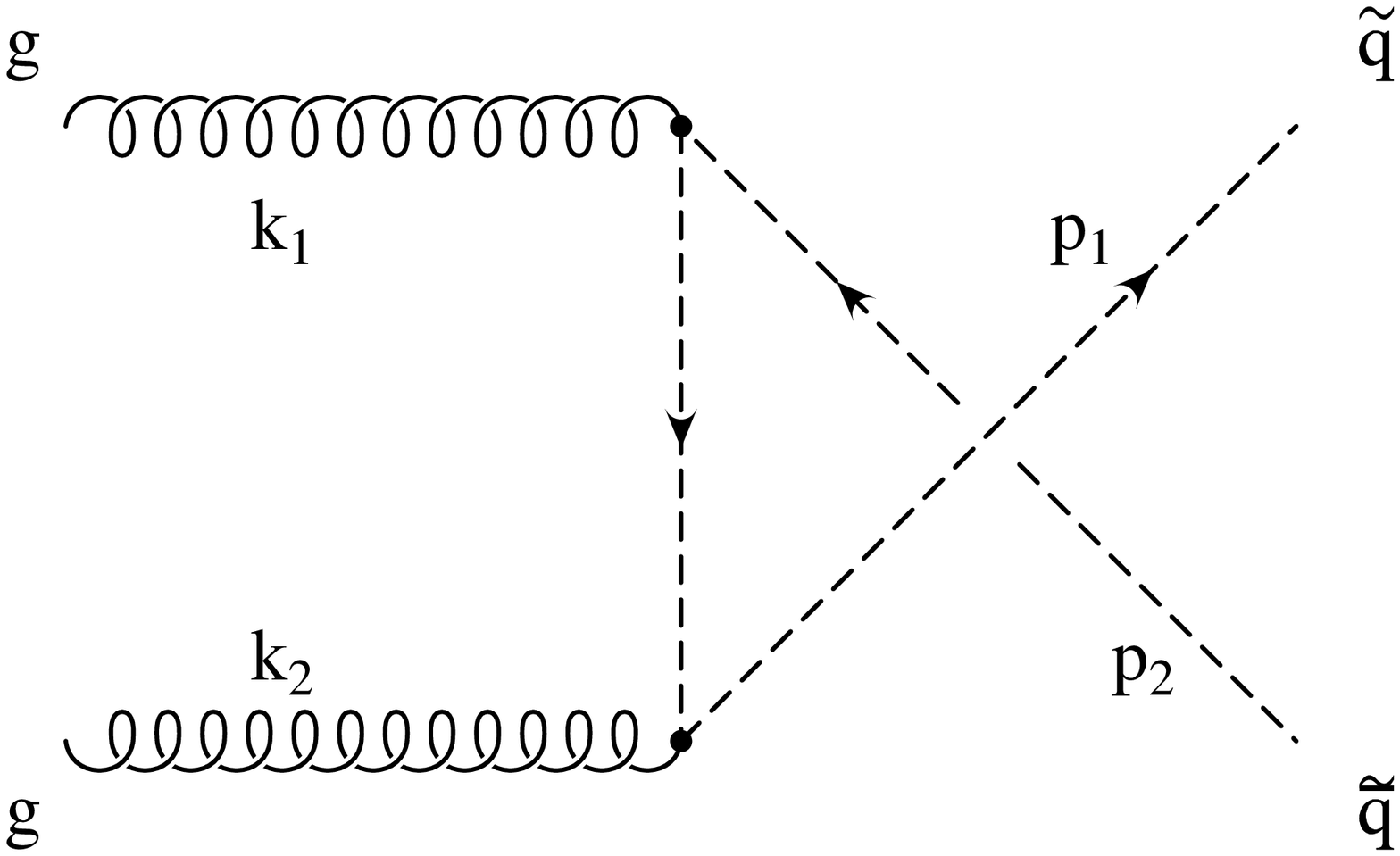,width=4cm}
\epsfig{file=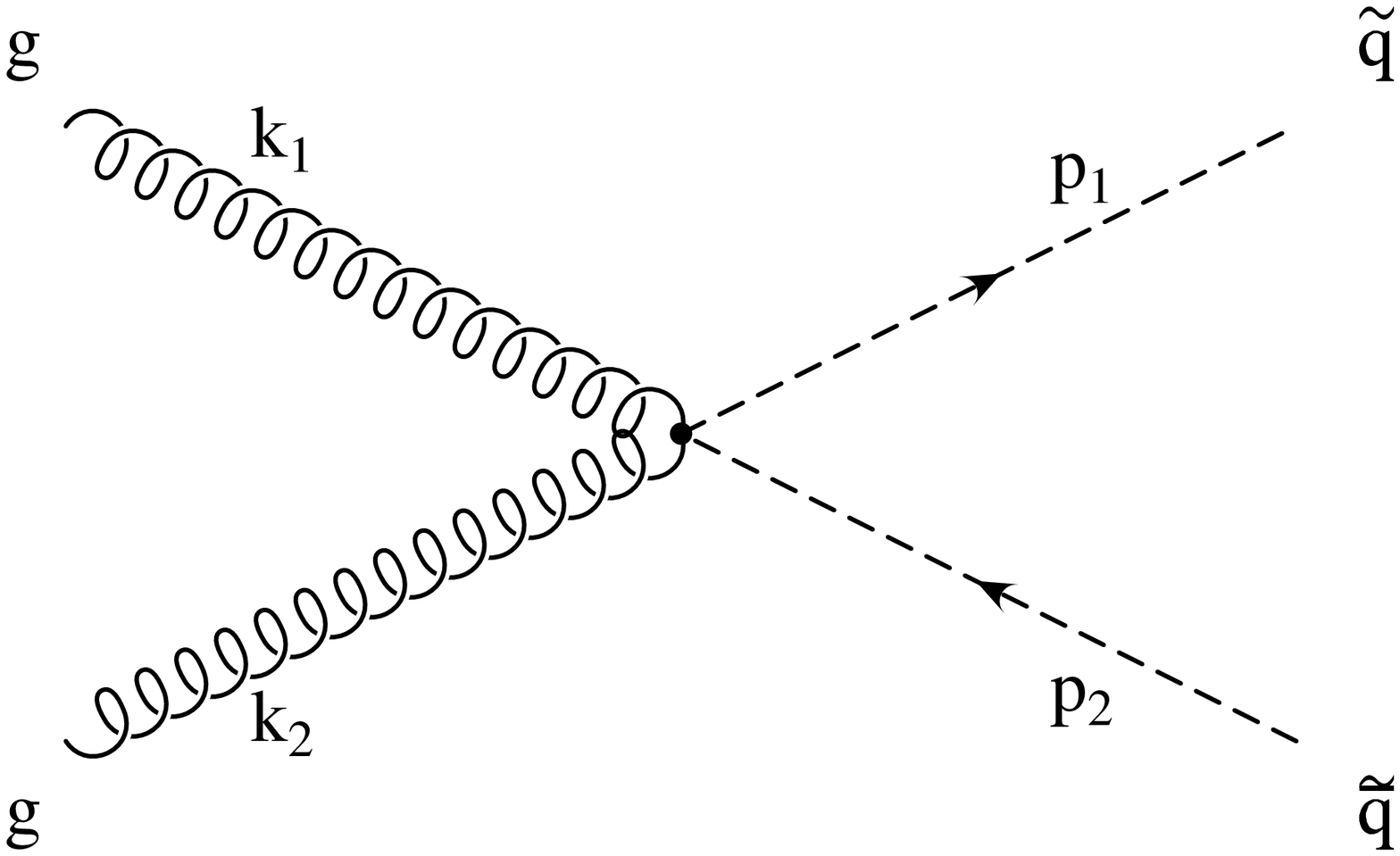,width=4cm}\\(b)
\end{center}
\caption{Feynman diagrams for squark-antisquark production: (a)
  quark-antiquark initial states, (b) gluon-gluon initial states.}
\label{fig:QQbar}
\begin{center}
\epsfig{file=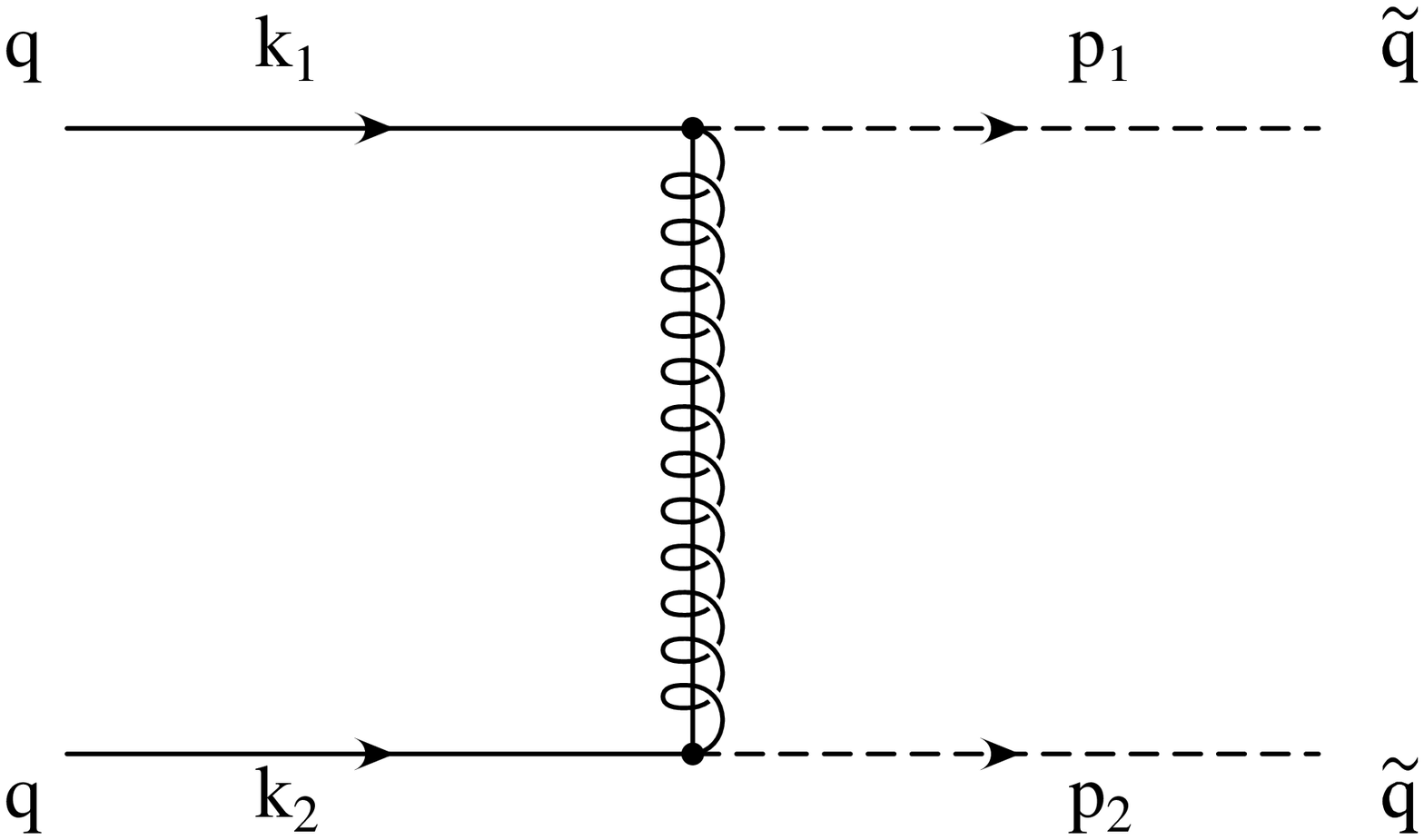,width=4cm}
\epsfig{file=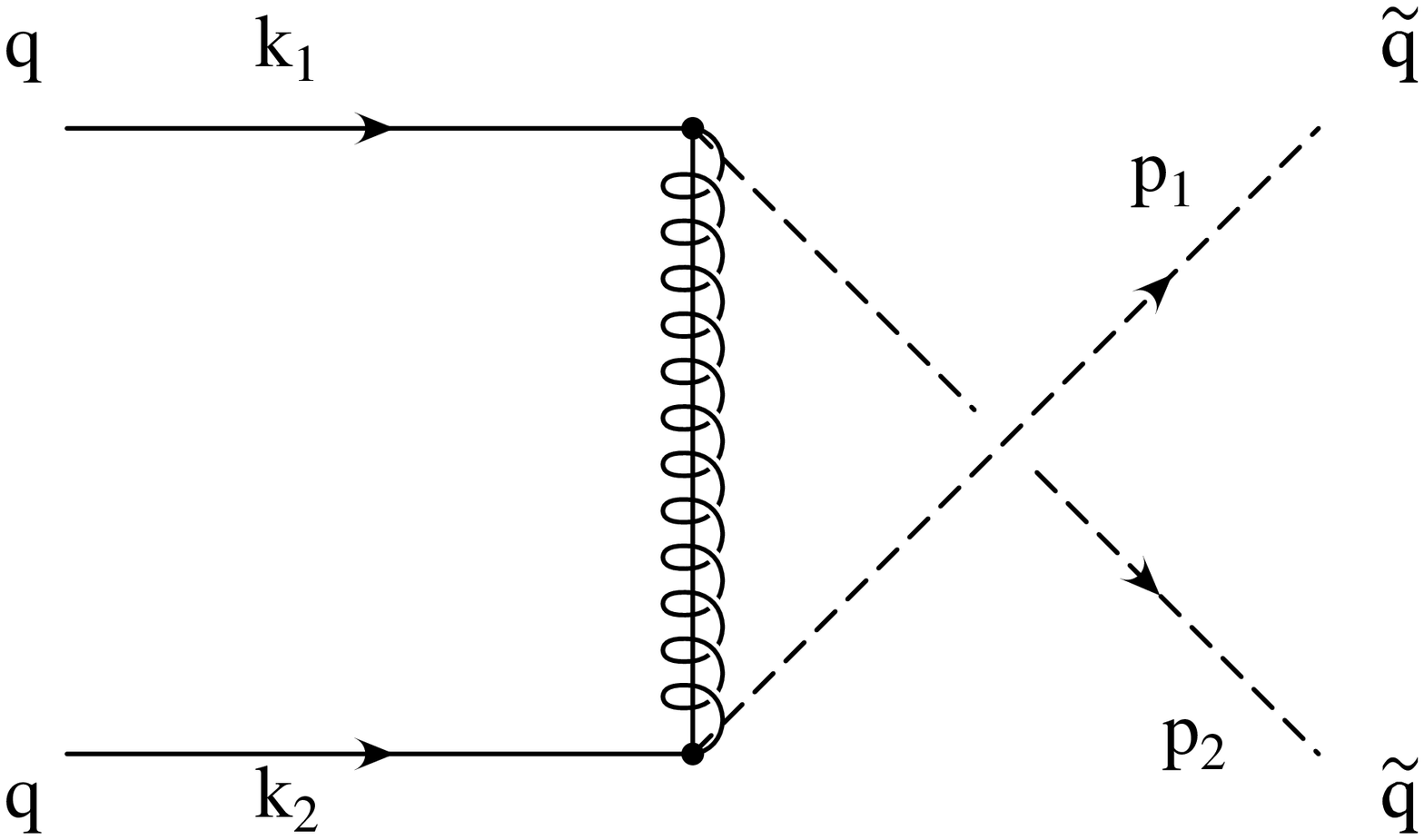,width=4cm}
\end{center}
\caption{Feynman diagrams for squark pair production.}
\label{fig:QQ}
\end{figure}

The Feynman diagrams for the 
partonic  squark-antisquark production processes 
in Figure~\ref{fig:QQbar}
give the squared matrix 
elements
\begin{eqnarray}
|{\cal M}|^2 (q_I+ \bar q_J \rightarrow \tilde q +\bar{\tilde
  q})\!\!&=&\!\!\delta_{IJ}g_s^4\left( 8n_f\frac{1}{s^2}NC_F(t_1
u_1-sm_{\tilde{q}}^2)-8\frac{1}{st_g}C_F(t_1
u_1-sm_{\tilde{q}}^2) \right)\nonumber \\ &{}& +4\frac{
  g_s^4}{t_g^2}NC_F\left(t_1 u_1-s(m_{\Q}^2 -m_{\G}^2)\right)\; , \\
\Delta |{\cal M}|^2 (q_I+ \bar q_J \rightarrow \tilde q +\bar{\tilde
  q})\!&\!=\!&\! - \delta_{IJ}g_s^4\left( 8n_f\frac{1}{s^2}NC_F(t_1
  u_1-sm_{\tilde{q}}^2)-8\frac{1}{st_g}C_F(t_1
  u_1-sm_{\tilde{q}}^2) \right)\nonumber \\ &{}&-4\frac{
  g_s^4}{t_g^2}NC_F\left(t_1 u_1-s \left(m_{\tilde{q}}^2
+m_{\tilde{g}}^2\right)\right)\;,\\
|{\cal M}|^2 (g+ g \rightarrow \tilde q +\bar{\tilde q})&=& 
4n_fg_s^4\left[C_O\left(1-2\frac{t_1u_1}{s^2}\right)-C_K\right]
\left[1-2\frac{sm_{\tilde
  q}^2}{t_1u_1}\left(1-\frac{sm_{\tilde q}^2}{t_1u_1}\right)\right]\;,
\\
\Delta |{\cal M}|^2 
(g+ g \rightarrow \tilde q +\bar{\tilde q})  &=& 4n_fg_s^4\left[C_O
\left(-1+2\frac{t_1u_1}{s^2}+2\frac{sm_{\Q}^2}{t_1u_1}-4
\frac{m_{\tilde q}^2}{s}\right)\right.\nonumber\\  
&{}&\left.\quad+C_K\left(1-2\frac{sm_{\tilde q}^2}{t_1u_1}
\right)\right]\;.\label{eq10}
\end{eqnarray}
%The unpolarized matrix elements are in agreement with~\cite{been}

The contributions to the 
partonic  squark-squark production are
shown in Figure~\ref{fig:QQ}. This process violates fermion number and occurs 
because of the Majorana nature of gluinos. We obtain
\begin{eqnarray}
|{\cal M}|^2(q+q\rightarrow \tilde q +\tilde q)
&=&\delta_{IJ} \left[2 g_s^4NC_F(t_1u_1-sm_{\tilde q}^2 )
\left(\frac{1}{t_g^2}+\frac{1}{u_g^2}\right) \right.\nonumber\\
&{}&\qquad +\left.  2 g_s^4m_{\tilde g}^2 s
\left(NC_F\left(\frac{1}{t_g^2}+\frac{1}{u_g^2}\right)
-2C_F\frac{1}{t_gu_g}\right)\right]\nonumber\\
&{}&+(1-\delta_{IJ})\left[4 g_s^4
NC_F\frac{t_1u_1-(m_{\tilde q}^2-m_{\tilde g}^2)s}{t_g^2}\right]\; \\
\Delta|{\cal M}|^2
(q+q\rightarrow \tilde q +\tilde q)&=&\delta_{IJ} 
\left[-2 g_s^4NC_F(t_1u_1-sm_{\tilde q}^2 )
\left(\frac{1}{t_g^2}+\frac{1}{u_g^2}\right)\right.\nonumber\\
&{}&\qquad \left.+2 g_s^4m_{\tilde g}^2s
\left(NC_F\left(\frac{1}{t_g^2}+\frac{1}{u_g^2}\right)
-2C_F\frac{1}{t_gu_g}\right)\right]\nonumber\\
&{}&+(1-\delta_{IJ})\left[4 g_s^4NC_F
\frac{-t_1u_1+(m_{\tilde q}^2+m_{\tilde g}^2)s}{t_g^2}\right]\;.
\label{eq12}
\end{eqnarray}
The second graph only contributes for final states of same flavour. 
We have already taken account of the symmetry factor 1/2 for the second 
term corresponding to the $LL$ or $RR$ initial quarks of same flavour, 
since in these cases the final states are indistinguishable.   
Our results for the unpolarized matrix elements agree with  
previous work~\cite{been}\footnote{There is a slight difference in the 
squared matrix elements because in ref.~\cite{been} the symmetry factors are
included at the level of the cross section.}. 
The polarized matrix elements (\ref{eq10}),
(\ref{eq12}) agree with~\cite{ratcliffe},
taking into account that we sum over final state squark chiralities
(and symmetrize where appropriate), while \cite{ratcliffe} presents
results for fixed squark chirality.

\begin{figure}[tb]
\begin{center}
\epsfig{file=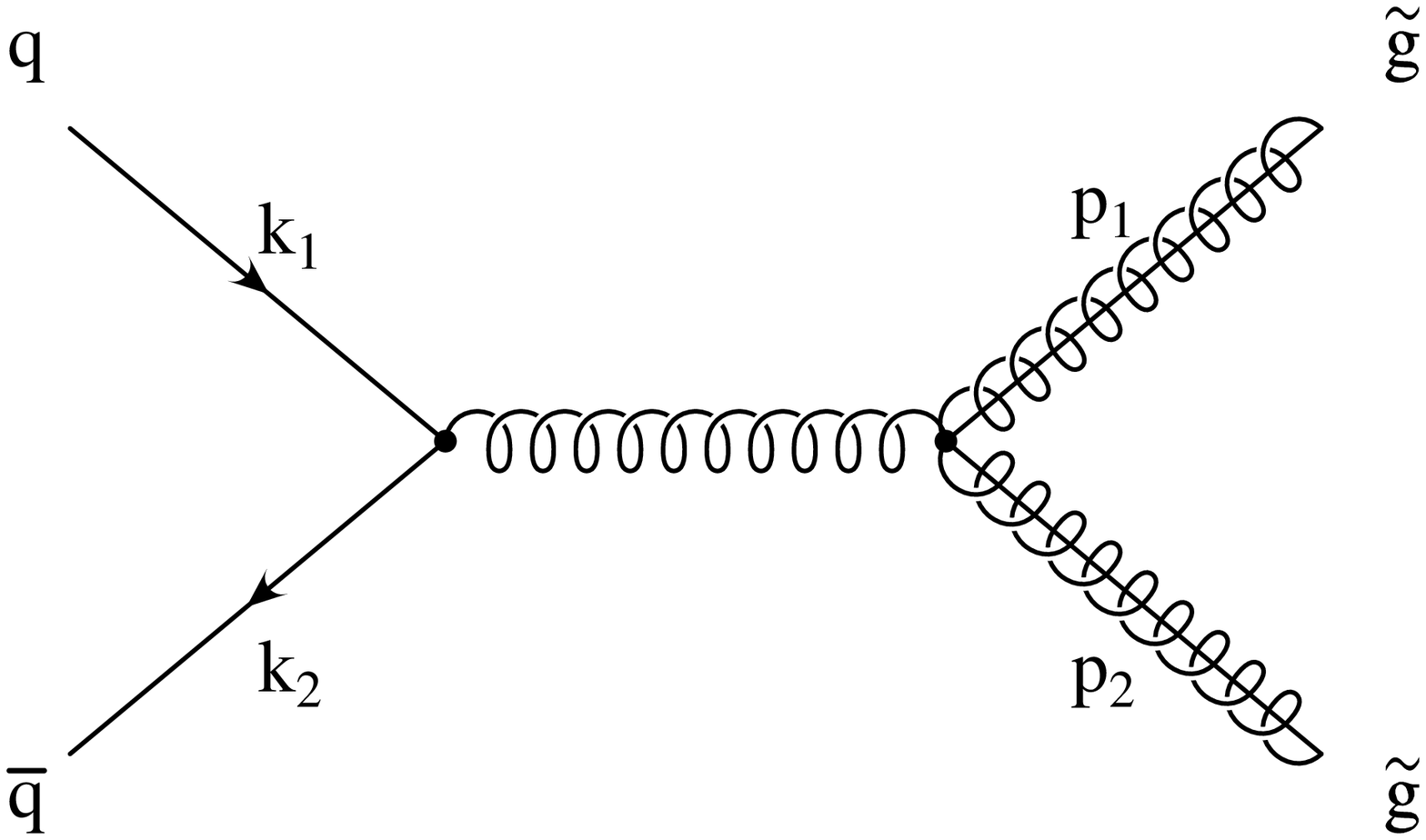,width=4cm}
\epsfig{file=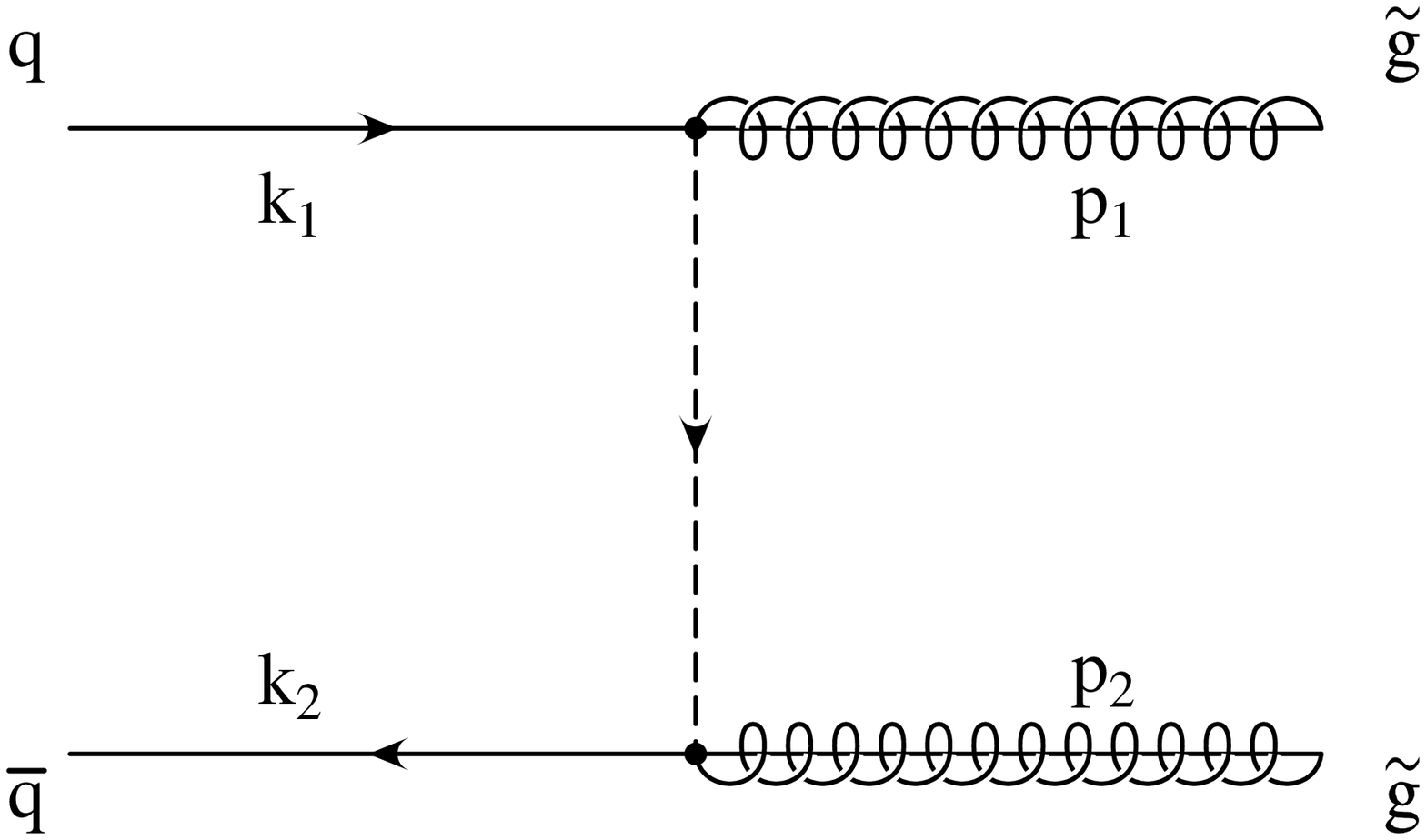,width=4cm}
\epsfig{file=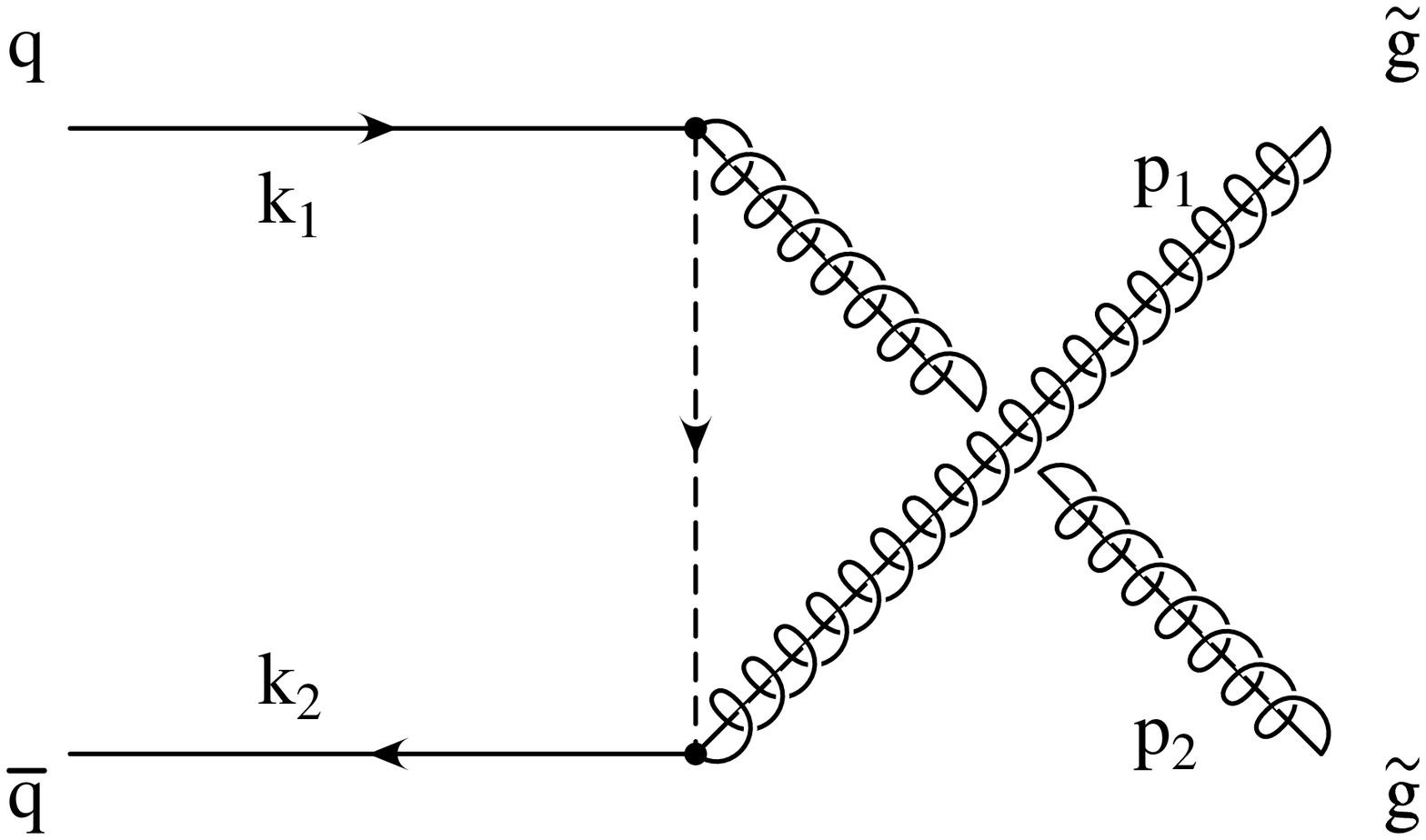,width=4cm}\\(a)\\
\epsfig{file=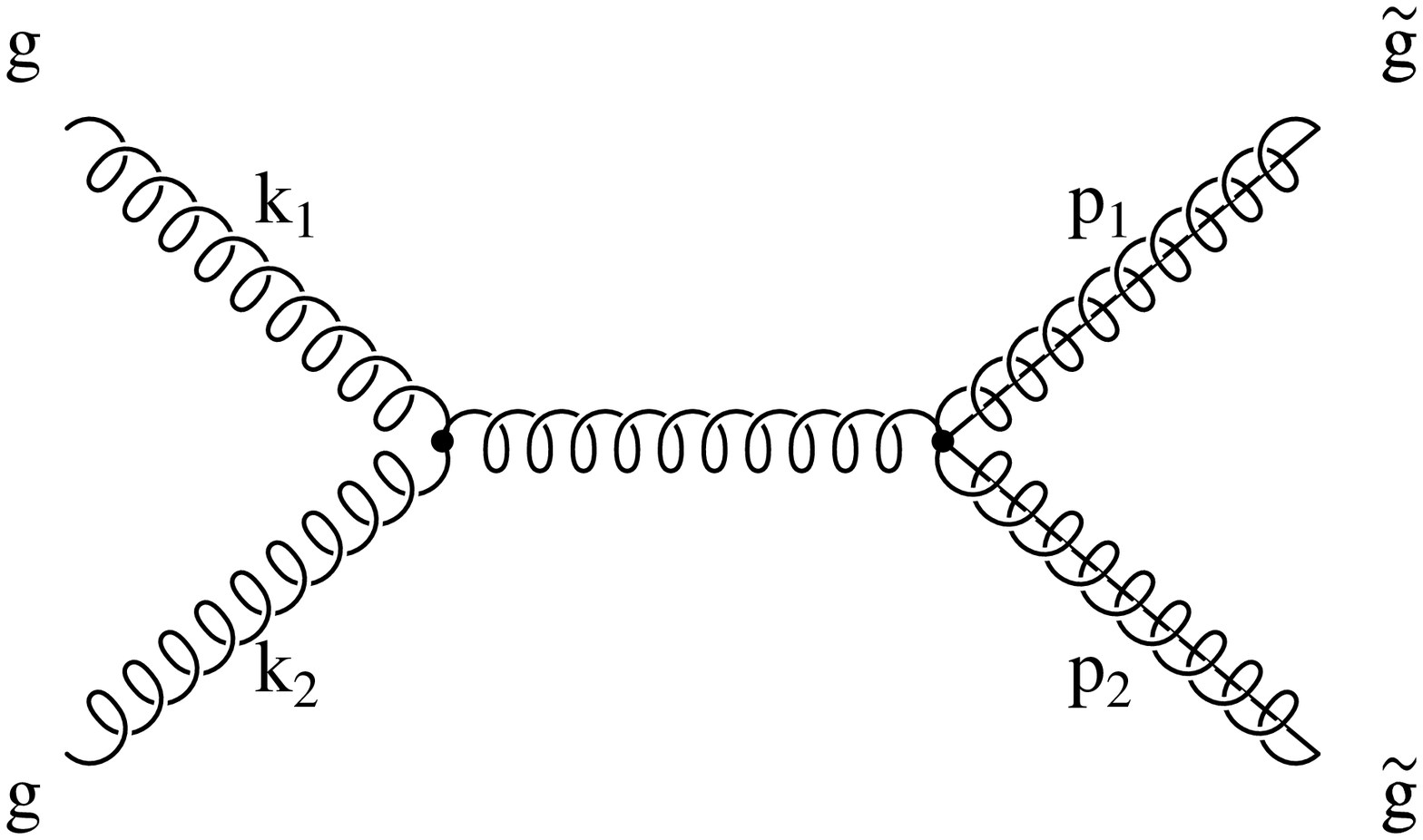,width=4cm}
\epsfig{file=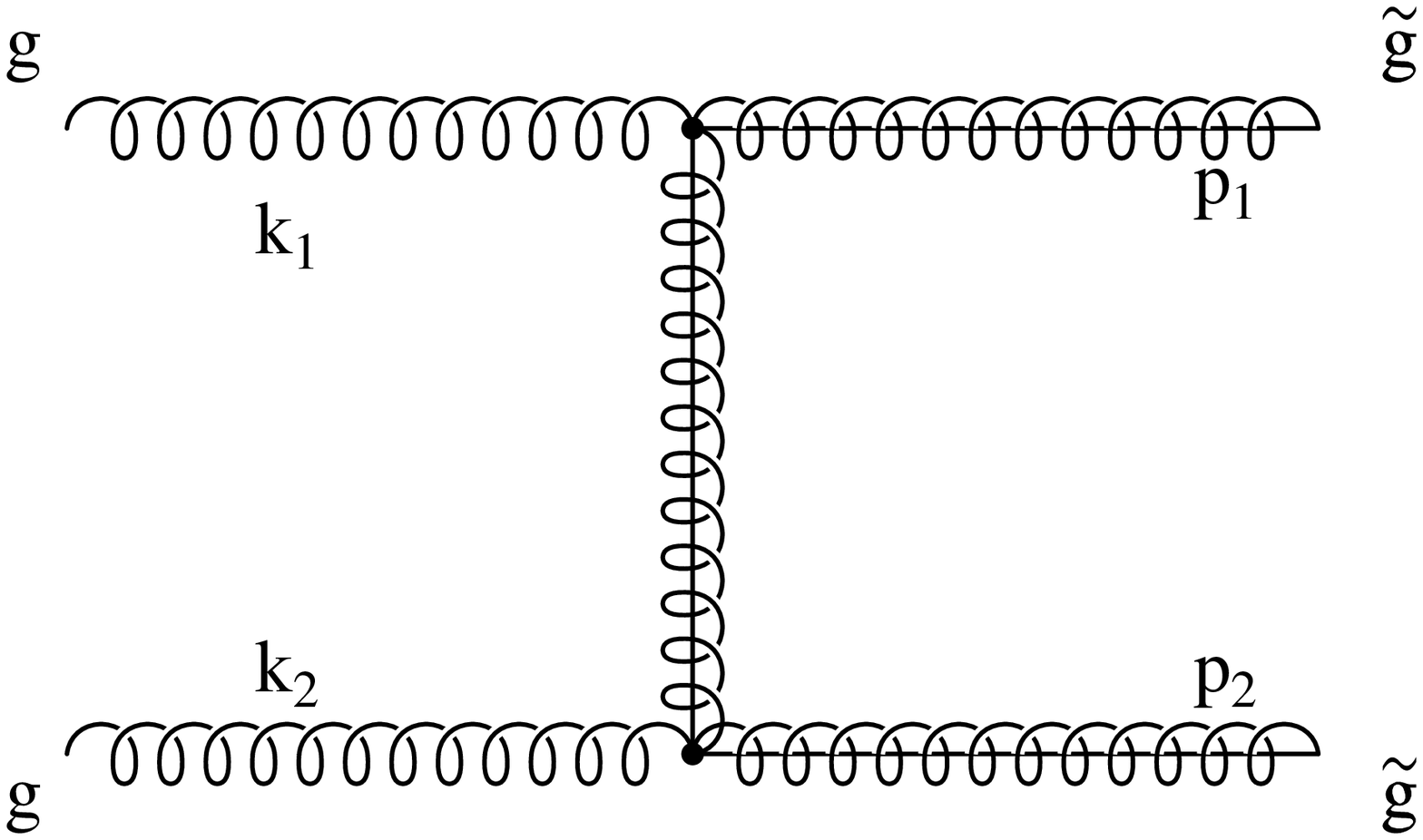,width=4cm}
\epsfig{file=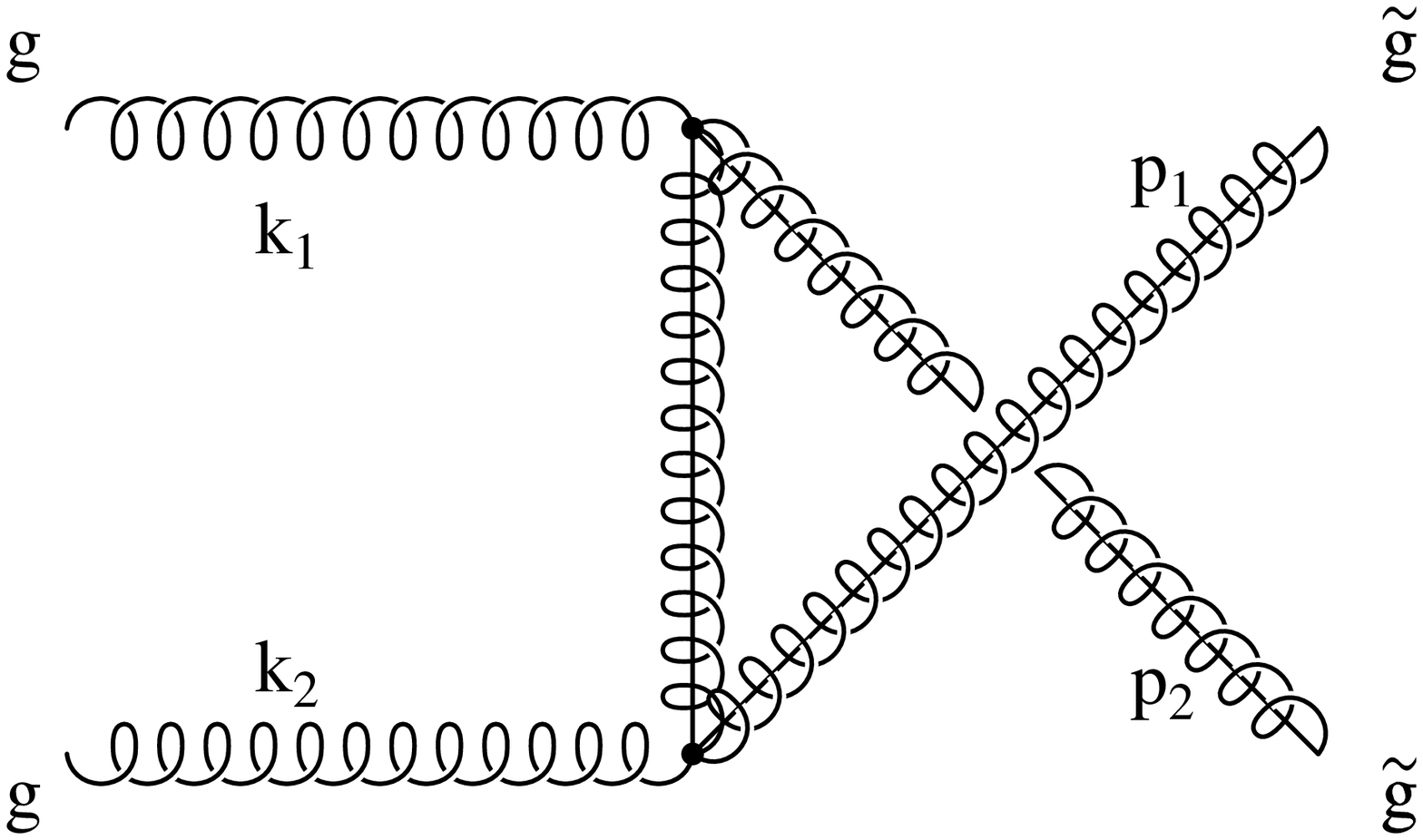,width=4cm}
\\(b)\\
\end{center}
\caption{Feynman diagrams for gluino pair production: (a)
  quark-antiquark initial states, (b) gluon-gluon initial states.}
\label{fig:GG}
\end{figure}
Gluino pairs can be produced in quark-antiquark or gluon-gluon annihilation as
shown in Figure~\ref{fig:GG}. 
For quark-antiquark initial states, only combinations with opposite helicities 
contribute and unpolarized and polarized matrix elements agree up to an
overall sign. The results are:
\begin{eqnarray}
|{\cal M}|^2 (q+\bar q \rightarrow \tilde g + \tilde g)&=
&2g_s^4C_O\left[\frac{2m_{\tilde g}^2s+t_g^2+u_g^2}{s^2}
+\frac{m_{\tilde g}^2s+t_g^2}{st_1}
+\frac{m_{\tilde g}^2s+u_g^2}{su_1}\right]\nonumber\\
&{}&\quad+g_s^4\left[C_O\left(\frac{t_g^2}{t_1^2}
+\frac{u_g^2}{u_1^2}\right)+C_K\left(2\frac{m_{\tilde g}^2s}{t_1u_1}
-\frac{t_g^2}{t_1^2}-\frac{u_g^2}{u_1^2}\right)\right]\; ,\\
\Delta |{\cal M}|^2 (q+\bar q \rightarrow \tilde g + \tilde g)&=
&- |{\cal M}|^2 (q+\bar q \rightarrow \tilde g + \tilde g)\; ,\\
|{\cal M}|^2(g+g\rightarrow \tilde g +\tilde g)&
=&4g_s^4 NC_O\left(1-\frac{t_g u_g}{s^2}\right)
\left[-2+\frac{s^2}{t_g u_g}+4\frac{sm_g^2 }{t_g u_g}
\left(1-\frac{sm_{\tilde g}^2 }{t_g u_g}\right)\right]\;,\\
\Delta|{\cal M}|^2(g+g\rightarrow \tilde g +\tilde g)&=&
4g_s^4NC_O\left[m_{\tilde g}^2\left(\frac{4}{s}
-\frac{6s}{t_gu_g}+2\frac{s^3}{u_g^2t_g^2}\right)
+3-2\frac{t_gu_g}{s^2}-\frac{s^2}{t_g u_g}\right].
\label{eq16}
\end{eqnarray}
For all gluino pair production matrix elements, we 
include a symmetry factor 
of $1/2$ to account for the production of identical Majorana 
fermions. Our results agree with ref.~\cite{been}, once the symmetry factor is 
taken into account.
\begin{figure}[tb]
\begin{center}
\epsfig{file=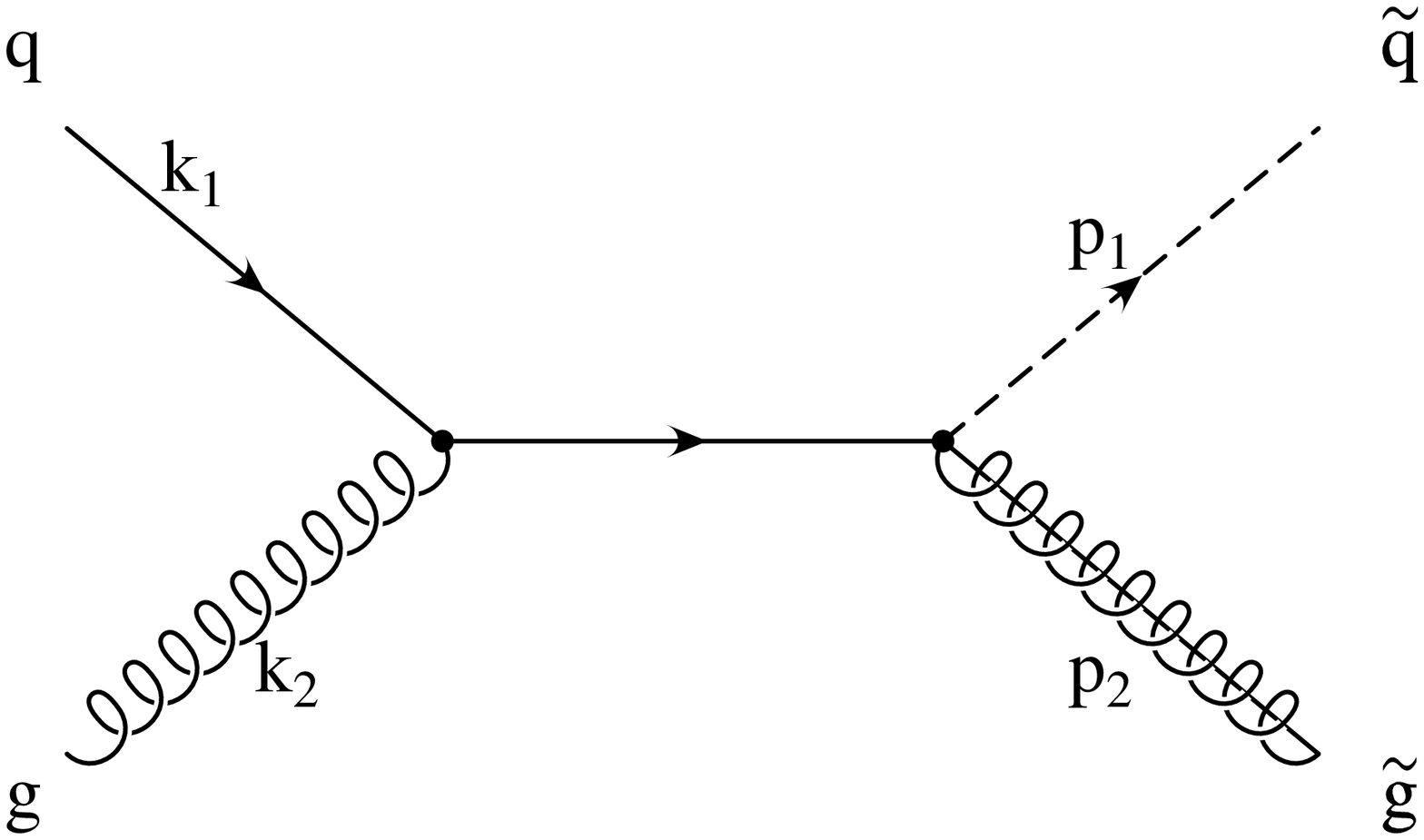,width=4cm}
\epsfig{file=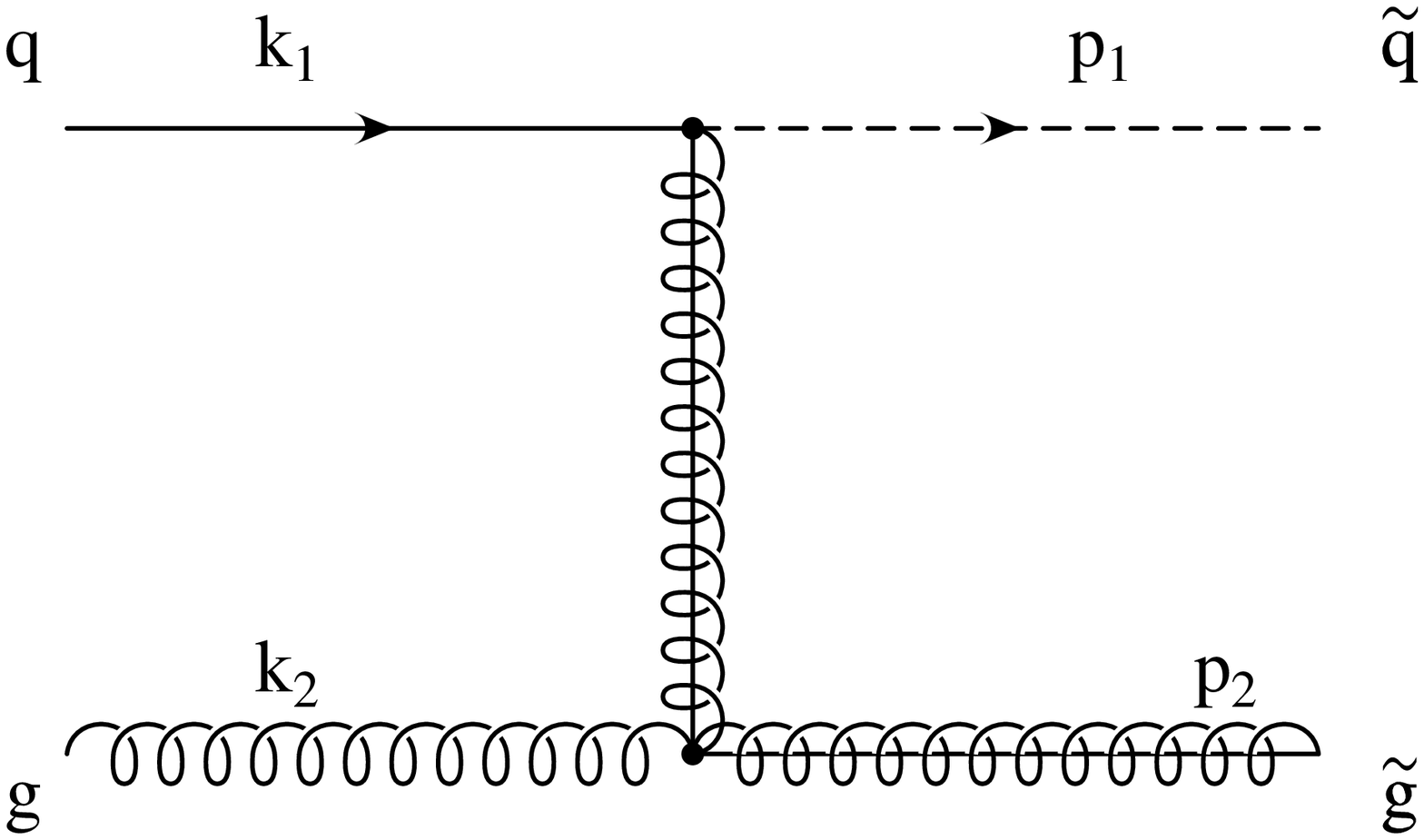,width=4cm}
\epsfig{file=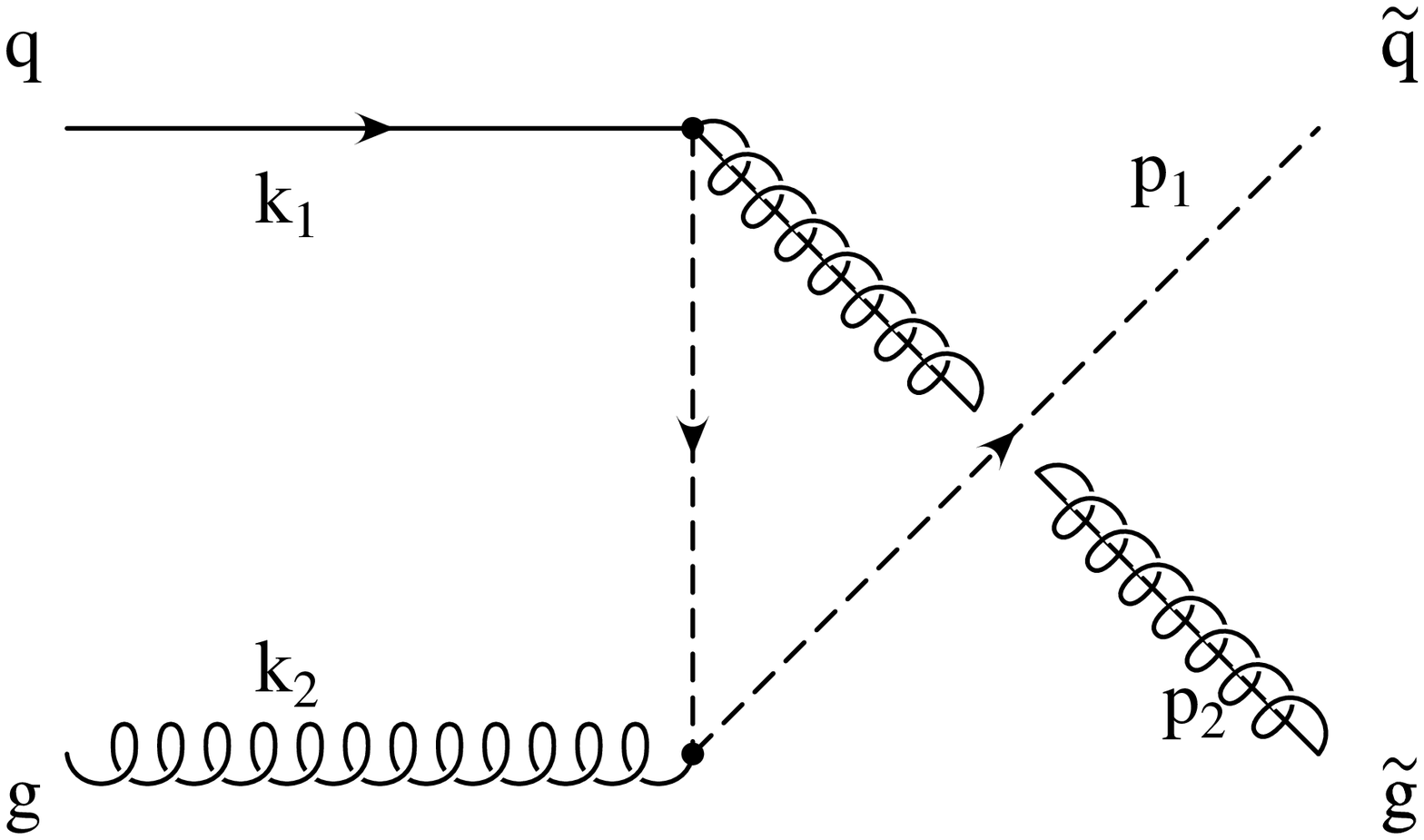,width=4cm}
\end{center}
\caption{Feynman diagrams for squark--gluino production.}
\label{fig:QG}
\end{figure}
Finally, squark-gluino final states are produced through the
quark-gluon scattering in Figure~\ref{fig:QG}, which yields 
\begin{eqnarray}
|{\cal M}|^2(q+g\rightarrow \tilde q + \tilde g )&=&
2g_s^4\left[C_O\left(1-2\frac{su_1}{t_g^2}\right)-C_K\right]
\nonumber \\ & &{}\times\left[-\frac{t_g}{s}+2\frac{(m_{\tilde g}^2
-m_{\tilde q}^2)t_g}{su_1}\left(1+\frac{m_{\tilde q}^2}{u_1}
+\frac{m_{\tilde g}^2}{t_g}\right)\right]\;,\\
\Delta|{\cal M}|^2(q+g\rightarrow \tilde q + \tilde g )&=&
2g_s^4\left[C_O\left[ m_{\tilde g}^2\left( 
4\frac{s}{t_g^2}+2\frac{s}{t_gu_1}+\frac{2}{t_g}
\right)\right.\right.\nonumber \\
& &{}\left.\left.-\left(2+\frac{t_g}{s}+2\frac{u_g}{s}
+2\frac{s}{t_g}+4\frac{u_g}{t_g}+2\frac{u_g}{u_1}\right)
\right]\right.\nonumber\\& &{}\left. +C_K\left(2\frac{m_{\tilde g}^2}{u_1}
-2\frac{t_gu_g}{su_1}+\frac{t_g}{s}\right)\right].
\label{eq18}
\end{eqnarray}
The unpolarized expressions agree with ref.~\cite{been}.
The polarized matrix elements (\ref{eq16}), (\ref{eq18}) 
agree with~\cite{ratcliffe},
again taking into account the different treatment of symmetry factors and
final state summations.

\subsection{Polarized partonic cross sections}
The inclusive cross section
for a particular partonic reaction where the two partons $i$ and $j$
in the initial state have helicities
$\lambda_i, \lambda_j$ is given by \footnote{We have omitted the labels $i,j$ for
the cross section where they are obvious.}
\begin{equation}
\hat{\sigma}_{\lambda_i, \lambda_j}
=\frac{K_{ij}}{2s}\int\!\! 
\frac{d^3p_1}{(2\pi)^32E_1}\frac{d^3p_2}{(2\pi)^32E_2}(2\pi)^4
\delta(k_1+k_2-p_1-p_2)|{\cal M}_{\lambda_i, \lambda_j}|^2.
\end{equation}
The factor $K_{ij}$ 
accounts for the average over the initial state colours. We have
$K_{q\bar q,qq}=1/N^2$ for quark-antiquark or quark-quark initial states, 
$K_{gg}=1/(N^2-1)^2$ for gluon-gluon initial states and $K_{qg}
=1/N/(N^2-1)$ 
four quark-gluon initial states. The above expression can be transformed into 
an integral over Mandelstam invariants
\begin{eqnarray}
\hat{\sigma}_{\lambda_i, \lambda_j}&=&K_{ij}\int\!\! 
du\!\!\int\!\! dt \frac{1}{16\pi s^2}
\Theta\left((t-m_2^2)(u-m_2^2)-sm_2^2\right)
\Theta(s-4m^2) \delta(s+t+u-m_1^2-m_2^2)|{\cal M}_{\lambda_i, \lambda_j}|^2
\nonumber \\
\end{eqnarray}
where $m=(m_1 + m_2)/2$ is the average mass of the final state particles.
For our goals we define the unpolarized and polarized cross sections by
\begin{eqnarray}
\hat{\sigma} &=&  \frac{1}{4}\left( \hat{\sigma}_{1,1} + \hat{\sigma}_{1,-1} + 
\hat{\sigma}_{-1,1} + \hat{\sigma}_{-1,-1} \right)\; ,\\
\Delta \hat{\sigma} &=&  \frac{1}{4}\left(
\hat{\sigma}_{1,1} - \hat{\sigma}_{1,-1} - 
\hat{\sigma}_{-1,1} + \hat{\sigma}_{-1,-1} \right) \; ,
\end{eqnarray}
where the symmetry factor $1/4$ accounts for 
an average over the partonic polarizations in the initial state.
%:
%$\hat{\sigma}_{\lambda_1,\lambda_2}$ denotes the cross section for 
%particular initial state polarizations. 

Carrying out the integration we obtain the following 
integrated cross sections which agree with \cite{been}:
\begin{eqnarray}
\hat{\sigma}(q_I+\bar q_J\rightarrow \tilde q + \bar{\tilde q})&=&\delta_{IJ}\frac{C_F}{N}\frac{n_f\pi\alpha_s^2}{s}\beta_{\tilde q}\left[\frac{1}{3}-\frac{4m_{\tilde q}^2}{3s}\right]\nonumber \\
&{}&+\delta_{IJ}\frac{C_F}{N^2}\frac{\pi\alpha_s^2}{s}\left[\beta_{\tilde q}\left(1+\frac{2m_{-}^2}{s}\right)+\left(\frac{2m_{\tilde g}^2}{s}+\frac{2m_{-}^4}{s^2} \right)L_1\right]\nonumber \\
&{}&+\frac{C_F}{N}\frac{\pi\alpha_s^2}{s}\left[\beta_{\tilde q}\left(-1-\frac{m_{-}^4}{m_{\tilde g}^2s+m_{-}^4}\right)+\left(-1-\frac{2m_{-}^2}{s} \right)L_1\right], \\
\hat{\sigma}(g+g \rightarrow \tilde q + \bar{\tilde q})&=&\frac{n_f\pi\alpha_s^2}{3(N^2-1)^2s^3}\left[C_O\left(\beta_{\tilde q}(2s^2+22sm_{\tilde q}^2)+12L_{\Q}sm_{\tilde q}^2\right)\right.\nonumber\\
&{}&\left.\qquad-C_K\left(\beta_{\tilde q}(3s^2+12sm_{\Q}^2)+12L_{\Q}sm_{\tilde q}^2-24L_{\Q}m_{\Q}^4\right)\right],\\
\hat{\sigma}(q_I+ q_J\rightarrow \tilde q + \tilde q)&=&\frac{C_F\pi\alpha_s^2}{Ns}\left[\beta_{\tilde q}\left(-1-\frac{m_{-}^4}{m_{\tilde g}^2s+m_{-}^4}\right)+\left(-1-\frac{2m_{-}^2}{s}\right)L_1\right]\nonumber\\
&{}&+\delta_{IJ}\frac{C_F}{N^2}\frac{\pi\alpha_s^2}{s}\left[\frac{2m_{\tilde
      g}^2}{s+2m_{-}^2}L_1\right],\\
\hat{\sigma}(q+\bar q\rightarrow \tilde g +\tilde g)&=&\frac{C_O}{N^2}\frac{\pi\alpha_s^2}{s}\beta_{\tilde g}\left(\frac{1}{3}+\frac{2m_{\tilde g}^2}{3s}\right)\nonumber \\
&{}&+\frac{C_O}{N^2}\frac{\pi\alpha_s^2}{s}\left[\beta_{\tilde g}\left(-\frac{1}{2}-\frac{m_{-}^2}{s}\right)+\left(\frac{m_{\tilde g}^2}{s}+\frac{m_{-}^4}{s^2} \right)L_2\right]\nonumber \\
&{}&+\frac{\pi\alpha_s^2}{s}\left[\frac{C_O-C_K}{N^2}\left(\beta_{\tilde g}\left(\frac{1}{2}+\frac{m_{-}^4}{2(m_{\tilde q}^2s+m_{-}^4)}\right)-L_2\frac{m_-^2}{s}\right)\right.
\nonumber\\
&{}& \qquad-\frac{C_K}{N^2}\left.\left(\frac{m_{\tilde g}^2}{s-2m_{-}^2} \right)L_2\right] ,\\
\hat{\sigma}(g+g \rightarrow \tilde g + \tilde g)&=&\frac{8NC_O\pi\alpha_s^2}{9(N^2-1)^2s}\left[\beta_{\tilde g}\left(-3-\frac{51m_{\tilde g}^2}{4s}\right)+\left(-\frac{9}{4}-\frac{9m_{\tilde g}^2}{s}+\frac{9m_{\tilde g}^4}{s^2}\right)L_{\G}\right],\\
\hat{\sigma}(q+g \rightarrow \tilde q + \tilde g)\!\!&\!\!=\!\!&\!\!-\frac{\pi\alpha_s^2}{4N(N^2-1)s^3}
	\left[
			C_K
			\left(
				4L_3(m_-^4-sm_-^2)+\kappa (s-7m_-^2)
			\right)\right.\nonumber \\
	&{}& 
			+C_O\left(
				  -4L_3(m_-^4-sm_-^2+2m_{\Q}^2m_-^2)\right.\nonumber \\
	&{}& \left.\left.+4L_4(2sm_-^2-2m_{\Q}^2m_-^2+s^2)+\kappa(3s+15m_-^2)
			\right)
	\right]		
	,
\end{eqnarray}
with the following functions and constants:
\begin{eqnarray}
\begin{array}{rclrcl}
\alpha_s&=&g_s^2/4\pi& &&\nonumber \\

\beta_{\Q}&=&\sqrt{1-\frac{4m_{\Q}^2}{s}} &
\beta_{\G}&=&\sqrt{1-\frac{4m_{\G}^2}{s}} \nonumber \\
m_{-}^2&=&m_{\G}^2-m_{\Q}^2 &
\kappa&=&\sqrt{(s-m_{\G}^2-m_{\Q}^2)^2-4m_{\G}^2m_{\Q}^2}\nonumber\\
L_1&=&\log\left(\frac{s+2m_{-}^2-s\beta_{\Q}}{s+2m_{-}^2+s\beta_{\Q}}\right) &
L_2&=&\log\left(\frac{s-2m_{-}^2-s\beta_{\G}}{s-2m_{-}^2+s\beta_{\G}}\right) \nonumber \\
L_3&=&\log\left(\frac{s-m_{-}^2-\kappa}{s-m_{-}^2+\kappa}\right) &
L_4&=&\log\left(\frac{s+m_{-}^2-\kappa}{s+m_{-}^2+\kappa}\right) \nonumber \\
L_{\Q}&=&\log\left(\frac{1-\beta_{\Q}}{1+\beta_{\Q}}\right) &
L_{\G}&=&\log\left(\frac{1-\beta_{\G}}{1+\beta_{\G}}\right) .
\end{array}\\
\end{eqnarray}

The new result concerns the polarized cross sections  $\Delta\hat{\sigma}$ 
which read:
\begin{eqnarray}
\Delta\hat{\sigma}(q_I+\bar q_J\rightarrow \tilde q + \bar{\tilde q})&=&-\delta_{IJ}\frac{C_F}{N}\frac{n_f\pi\alpha_s^2}{s}\beta_{\tilde q}\left[\frac{1}{3}-\frac{4m_{\tilde q}^2}{3s}\right]\nonumber \\
&{}&-\delta_{IJ}\frac{C_F}{N^2}\frac{\pi\alpha_s^2}{s}\left[\beta_{\tilde q}\left(1+\frac{2m_{-}^2}{s}\right)+\left(\frac{2m_{\tilde g}^2}{s}+\frac{2m_{-}^4}{s^2} \right)L_1\right]\nonumber \\
&{}&-\frac{C_F}{N}\frac{\pi\alpha_s^2}{s}\left[\beta_{\tilde q}\left(-3-\frac{m_{-}^4}{m_{\tilde g}^2s+m_{-}^4}\right)+\left(-1-\frac{2m_{-}^2}{s} \right)L_1\right], \\
\Delta\hat{\sigma}(g+g \rightarrow \tilde q + \bar{\tilde
  q})&=&-\frac{n_f\pi\alpha_s^2}{3\left(N^2-1\right)^2 s^3}
\left[
C_O\left(
	\beta_{\tilde q}(2s^2+10sm_{\tilde q}^2)+12L_{\Q}sm_{\tilde q}^2
	\right)\right.\nonumber\\ &{}&\left.	
+C_K\right(
	-3\beta_{\Q}s^2-12sm_{\tilde q}^2L_{\Q}
	\left)
\right],\\
\Delta\hat{\sigma}(q_I+ q_J\rightarrow \tilde q + \tilde q)&=&\frac{C_F\pi\alpha_s^2}{Ns}\left[\beta_{\tilde q}\left(3-\frac{m_{-}^4}{m_{\tilde g}^2s+m_{-}^4}\right)+\left(1+\frac{2m_{-}^2}{s}\right)L_1\right]\nonumber\\
&{}&+\delta_{IJ}\frac{C_F}{N^2}\frac{\pi \alpha_s^2}{s}\left[\frac{2m_{\tilde g}^2}{s+2m_{-}^2}L_1\right],\\
\Delta\hat{\sigma}(q+\bar q\rightarrow \tilde g+ \tilde g)&=&-\hat{\sigma}(q\bar q\rightarrow \tilde g \tilde g),\\
\Delta\hat{\sigma}(g+g \rightarrow \tilde g + \tilde g)&=&\frac{NC_O}{\left(N^2-1\right)^2}\frac{\pi\alpha_s^2}{s}\left[\beta_{\tilde g}\left(\frac{20}{3}+\frac{10m_{\tilde g}^2}{3s}\right)+\left(2+\frac{4m_{\tilde g}^2}{s}\right)L_{\G}\right],\\
\Delta\hat{\sigma}(q+g \rightarrow \tilde q + \tilde g)&=&\frac{\pi\alpha_s^2}{4N(N^2-1)s^3}
	\left[
		C_K
		\left(
			4L_3sm_{\Q}^2+\kappa(s-3m_{-}^2)
		\right)
		\right.\nonumber\\ &{}&\left.+C_O
		\left(
			-4L_3sm_{\Q}^2+4L_4s(s+2m_{-}^2)+\kappa(11s+3m_{-}^2)
		\right)
	\right].
\end{eqnarray}

In all cross sections, the production threshold factor $\Theta(s -
 4 m^2)$ is implicit. 

\subsection{Polarized hadronic cross sections}
The measured hadronic cross sections $\sigma(A+B\to X)$ are obtained by  
weighting the
partonic cross sections $\hat \sigma_{\lambda_1,\lambda_2}$ 
with the probability of finding the initial partons with the corresponding
helicity orientations inside the initial hadrons $A$ and $B$ which are assumed
to be polarized.
We are particularly interested in the polarized 
cross section
$\Delta \hat \sigma$. It is useful to define the unpolarized and polarized 
cross sections by  
\begin{eqnarray}
\sigma(A+B\to X) &=& \sum_{
\textnormal{process } \atop i+j\rightarrow X}
\int_0^1 \d x_1 \d x_2 
\Big(p_{i/A}(x_1)p_{j/B}(x_2) 
\label{eq:sig}
\nonumber \\&& \hspace{2cm}
+ p_{j/A}(x_1)p_{i/B}(x_2) \Big)
\hat\sigma_{i+j\rightarrow X}(s)\;, \\
\Delta \sigma(A+B\to X) &=& \sum_{
\textnormal{process } \atop i+j\rightarrow X}
\int_0^1 \d x_1 \d x_2 
\Big(\Delta p_{i/A}(x_1)\Delta p_{j/B}(x_2) \nonumber \\&& \hspace{2cm}
+ \Delta p_{j/A}(x_1)\Delta p_{i/B}(x_2) \Big)
\Delta \hat\sigma_{i+j\rightarrow X}(s)\;,
\label{eq:delsig}.
\end{eqnarray}
$S$ is the centre-of-mass energy squared at the hadronic and
$s=Sx_1x_2$ at the partonic level. $p_{k/H}(x)$ and $\Delta p_{k/H}(x)$ 
are the unpolarized and polarized parton distribution functions. The unpolarized
distributions parametrise the probability of finding a parton $k$ 
inside a hadron $H$ with momentum 
fraction $x$ and any helicity, whereas the the polarized distributions
give the difference between the probabilities of finding a parton with its 
spin parallel and anti-parallel to that of the hadron. The summation 
includes all quark and antiquark flavours and the gluon. 

The above hadronic cross sections can be measured experimentally by 
studying production cross sections with different
but fixed longitudinal spin-orientations 
for the incoming hadrons $\sigma^{\pm\pm}$, where each $(\pm)$ denotes the 
longitudinal spin orientation (in direction of motion or opposite) 
of one of the incoming hadrons:
\begin{eqnarray}
\sigma &=& \frac{1}{4} \left(\sigma^{++} + \sigma^{+-} + \sigma^{-+} 
+ \sigma^{--} \right)\; ,\\
\Delta \sigma &=& \frac{1}{4} \left(\sigma^{++} - \sigma^{+-} - \sigma^{-+} 
+ \sigma^{--} \right)\; .
\end{eqnarray}
From these, one can construct the production asymmetry
\begin{equation}
A = \frac{\Delta \sigma}{\sigma}\; .
\label{asymmetry}
\end{equation}
Because several uncertainties in the normalisation cancel,
$A$ can be measured with less systematic errors than the 
individual cross sections. 
\section{Inclusive hadronic asymmetries}
\label{sec:num}

To illustrate the potential of polarized hadron colliders, we consider
%Numerically evaluating (\ref{eq:sig}) and (\ref{eq:delsig}) for fixed $S$
%and different collider types (proton-proton or 
%proton-antiproton), we calculate expected production 
%cross sections and spin asymmetries. 
the following three possibilities:

\noindent\parbox{16cm}{\hspace{0.9cm}
\parbox{15cm}{
\begin{itemize}
\item[{{\bf RHIC:}}] At present, the only polarized hadron collider is 
the relativistic heavy ion collider (RHIC) at BNL, which collides 
two longitudinally polarized proton beams at centre-of-mass energies 
$\sqrt{S} = 200$ to  $500$~GeV. We consider only $\sqrt{S} = 500~$GeV, 
with a beam polarization 
of 70\% for both beams and an integrated luminosity of 800 pb$^{-1}$, 
according to the design parameters~\cite{rhic}. 
\item[{{\bf Tevatron:}}] The Fermilab Tevatron is an unpolarized 
proton-antiproton collider, which is currently operating at 
$\sqrt{S} = 1.96$~TeV. In the current run, an integrated luminosity of 
10 fb$^{-1}$ is anticipated. We take these values and a hypothetical 
%To illustrate the physics potential of 
%operating the Tevatron with polarized beams, we
%compute cross sections and asymmetries for this 
%proton-antiproton centre-of-mass energy and luminosity, assuming moreover
70\% polarization for both beams. 
\item[{{\bf LHC:}}] At present under construction at CERN, the 
large hadron collider will soon provide proton-proton collisions at 
$\sqrt{S} = 14$~TeV. To examine the physics prospects of a 
 hypothetical polarized physics programme, we 
assume an integrated luminosity of 500 fb$^{-1}$ and beam polarizations of 
70 \%. 
\end{itemize}
}}

We numerically integrated the formulae for the total cross section 
(\ref{eq:sig}) and for the 
hadronic polarisation difference (\ref{eq:delsig}) 
to calculate the asymmetry $A(Z)$ (formula (\ref{asymmetry})) 
for all squark and gluino final states 
$X=\G\G, \G\Q/\G\bar\Q, \Q\Q/\bar\Q\bar\Q, \Q\bar\Q$. 
The GRV98 unpolarized parton density parametrisation  \cite{GRV98} 
for the $u,d,s$ quarks was used. These distributions are determined in 
a fixed flavour 
number scheme~\cite{hqs}, and thus do not contain $c$ and $b$ quarks. As
in the unpolarized calculation of~\cite{been}, 
we take these from the the older 
GRV92 parametrisation~\cite{GRV92}. 
For the polarised 
parton distributions, we used the  
GRSV2000 parametrisation~\cite{grsv} with its
"standard" scenario, which contains only $u,d,s$ quarks. The polarized heavy 
quark distributions are expected to be negligible compared to unpolarized
ones. We use throughout the leading order densities as well as the
leading order strong coupling constant with $\lambda_{QCD}^{n_f=5} = 
132$~MeV ($\alpha_s^{LO}(M_Z) = 0.125$). 
\begin{figure}[t]
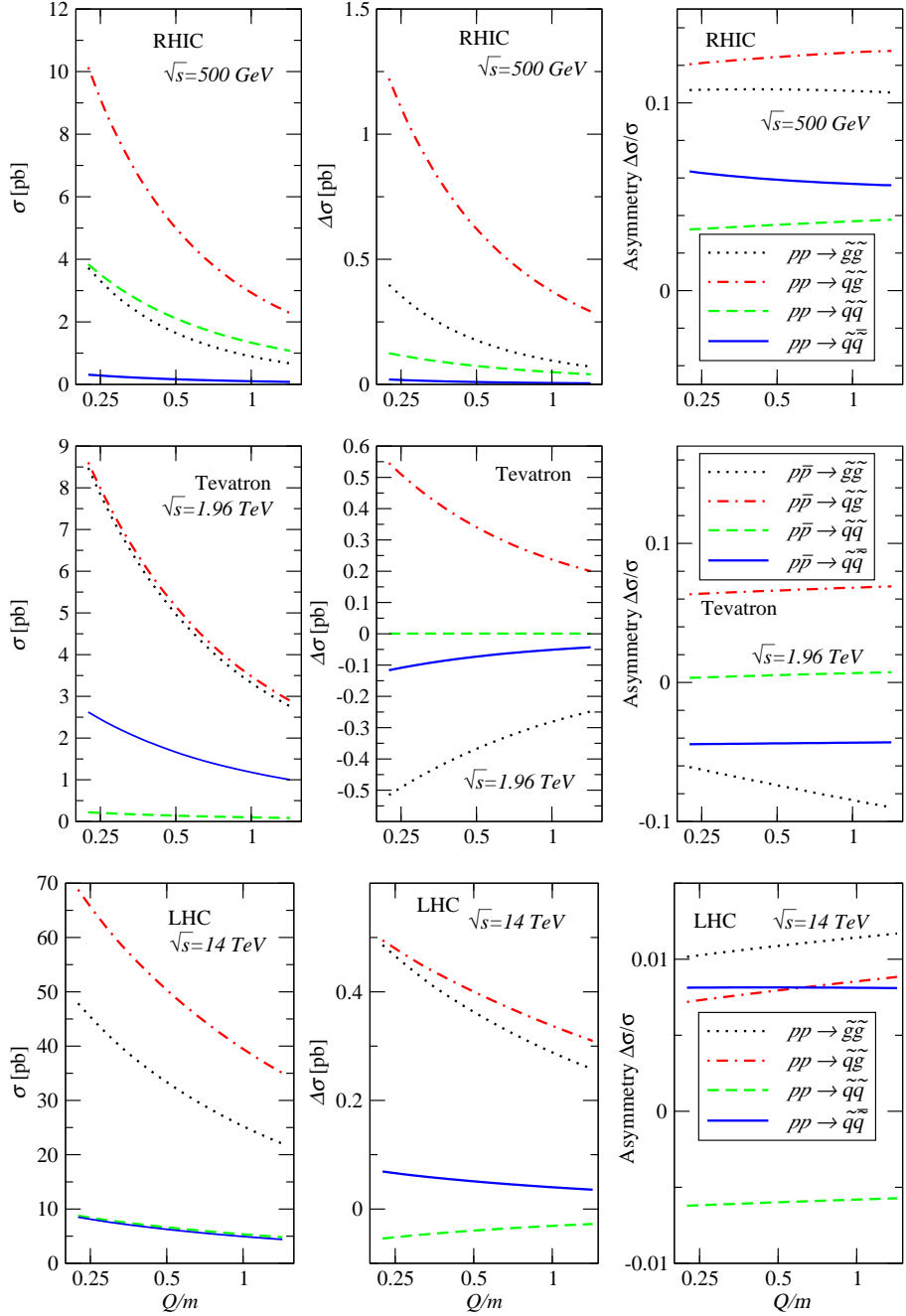

	\begin{center}
\epsfig{file=scale_sum_diff_rhic.eps,width=12cm}
\epsfig{file=scale_sum_diff_Teva.eps,width=12cm}
\epsfig{file=scale_sum_diff_lhc.eps,width=12cm}
\caption{Dependence of total unpolarized and polarized production cross 
sections and asymmetry on the renormalisation and 
mass factorisation scale $Q$ for 
RHIC, Tevatron and LHC. The mass parameters are for RHIC: $m_{\Q}=100$ GeV,
 $m_{\G}=80$ GeV; Tevatron: $m_{\Q}=280$ GeV, $m_{\G}=200$ GeV and 
LHC: $m_{\Q}=600$ GeV, $m_{\G}=500$ GeV. 
$Q$ is the renormalisation and factorisation scale 
for the parton distributions and $m=(m_1+m_2)/2$, 
where $m_1$ and $m_2$ are the mass of the produced particles.}
	\end{center}
\label{asym-scale-lhc}
\end{figure}

\subsection{Scale dependence}
Both the QCD coupling constant $\alpha_s$ and the parton distribution functions
depend on the scale at which they are probed. Working at leading order
 the production cross sections are therefore scale-dependent. 
Since polarized and unpolarized cross sections scale with the same power of
$\alpha_s$, the renormalisation scale dependence drops out in the asymmetry.
The remaining dependence on the factorisation scale of the 
parton distribution functions is weak because the evolution of the 
distribution functions in the numerator and denominator 
are comparable. This is illustrated in Figure 
\ref{asym-scale-lhc} for the three 
colliders: the scale dependence of the asymmetry almost disappears.
%While the sca
%The first two graphs of each figure show the scale dependence 
%of the total cross section and the polarisation 
%difference. The last graph shows the scale dependence of the asymmetry.
%It can clearly be seen that the dependence
%on the renormalization and factorization scale is much weaker 
%for  the spin asymmetry 
%than for unpolarized and polarized cross sections, as expected from the 
%above arguments. 
This could be taken as an indication that the leading order 
spin asymmetry does not receive substantial corrections in higher order
and is sufficient for our exploratory studies. We will set the renormalisation
and factorisation scale to the average final state mass $m$. 

\subsection{Asymmetries at different colliders}
\begin{figure}[t]
	\begin{center}
		\epsfig{file=s_asym_all.eps,width=12cm}
	\end{center}
	\caption{Asymmetry as a function of the hadronic centre of mass energy.}
\label{s_asym_all}
	\begin{center}
		\epsfig{file=s_sep_QQ.eps,width=12cm}
	\end{center}
	\caption{Quark-antiquark and gluon pair contributions to asymmetry of the squark-antisquark production as a function of the hadronic centre of mass energy.}
\label{s_sep_QQbar}
\end{figure}
\begin{figure}[t]
	\begin{center}
		\epsfig{file=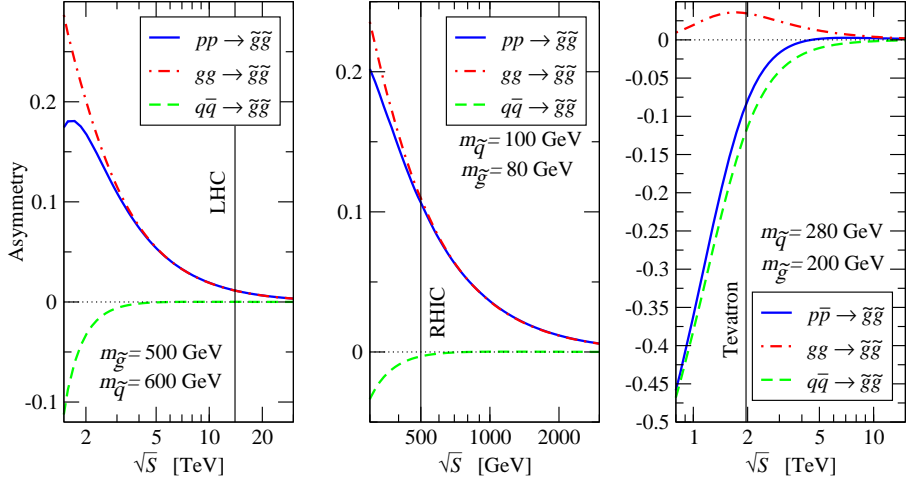,width=12cm}
	\end{center}
	\caption{Quark-antiquark and gluon pair contributions 
to the asymmetry for the gluino pair production as a function of the 
hadronic centre of mass energy.}
\label{s_sep_GG}
\end{figure}
 
Figure~\ref{s_asym_all} shows the asymmetries for different colliders
and centre-of-mass energies for fixed squark and gluino masses.
%collider energy, we study the behaviour of the asymmetry for 
%fixed squark and gluino masses as function of the collider centre-of-mass 
%energy, as displayed in Figure~\ref{s_asym_all}, where we also indicated the 
%parameters corresponding to the three collider scenarios we consider. 
For the masses chosen, the  asymmetries at LHC and RHIC are quite similar; 
only the signs in squark-squark production are different as a consequence
of the rapid falloff of the polarisation difference with $S$.
In contrast, the asymmetries at
proton-antiproton colliders are quite different; 
both the gluino-gluino and the squark-antisquark 
asymmetries are negative. 

These two processes obtain 
contributions from the quark-antiquark 
initial state and  from the gluon pair initial state.
The relative contributions to the asymmetry from the quark-antiquark 
and gluon pair initial states are illustrated in more detail 
Figures \ref{s_sep_QQbar} and \ref{s_sep_GG}. 
For squark-antisquark production, 
Figure \ref{s_sep_QQbar},
we find that for $pp$ machines (LHC and RHIC), 
the quark-antiquark initial state is more important for $\sqrt{S}$ below
four times the squark mass, whereas at higher $\sqrt{S}$
gluon pairs dominate.
In $p\bar p$ collisions, on the other hand, the quark-antiquark initial state
dominates always (except where its contribution changes sign --
at $\sqrt{S}\simeq 1,2$ TeV in our case).
For the 
gluino pair production, Figure  \ref{s_sep_GG}, 
the gluon initial state (positive contribution) 
dominates in $pp$ collisions (LHC and RHIC),
and the quark-antiquark initial state (negative contribution)
in $p\bar p$ collisions. 
At LHC, the latter contribution 
turns positive because the polarised density product 
$\Delta q(x)\Delta \bar q(x)$ changes sign in the neighbourhood 
of $x=0.2$, which is only accessible at the LHC  
with our choice of mass parameters.

The asymmetries are largest in the neighbourhood of the production threshold, 
$\sqrt{S}/2 \simeq m$ where particle production 
requires $x$ to be near $1$. Since the parton distributions 
are steeply falling and 
badly known for $x\rightarrow 1$, these apparently large asymmetries 
correspond to vanishing cross sections and cannot be reliably predicted.

Figures~\ref{s_sep_QQbar} 
and~\ref{s_sep_GG} also illustrate 
the uncertainty in the predicted asymmetry due to the
uncertainties in the (polarized) parton distribution functions.
Most sparticle production processes (except squark--squark 
production) at LHC are dominated by gluon initial states. 
Since the 
polarized gluon distribution is not well constrained by experimental data 
at present, one should consider the predictions for the production 
asymmetries at LHC as uncertain. 
The  present fits of polarized parton distributions come largely from polarized 
deep inelastic scattering data which probe only the polarized quark 
distributions. The polarized gluon distributions are obtained from the 
evolution of the polarized structure functions, and are therefore only weakly 
constrained. Thus the predicted asymmetries are less reliable 
in energy and mass ranges where initial gluons are important.
However, data on single inclusive 
hadron production and on  jet production from RHIC will soon improve 
our knowledge on the polarized gluon distribution, such that more 
accurate predictions will become feasible. 
On the other hand, squark-squark production asymmetries which only requires
quark-quark initial states are already quite accurate.
\begin{figure}[t]
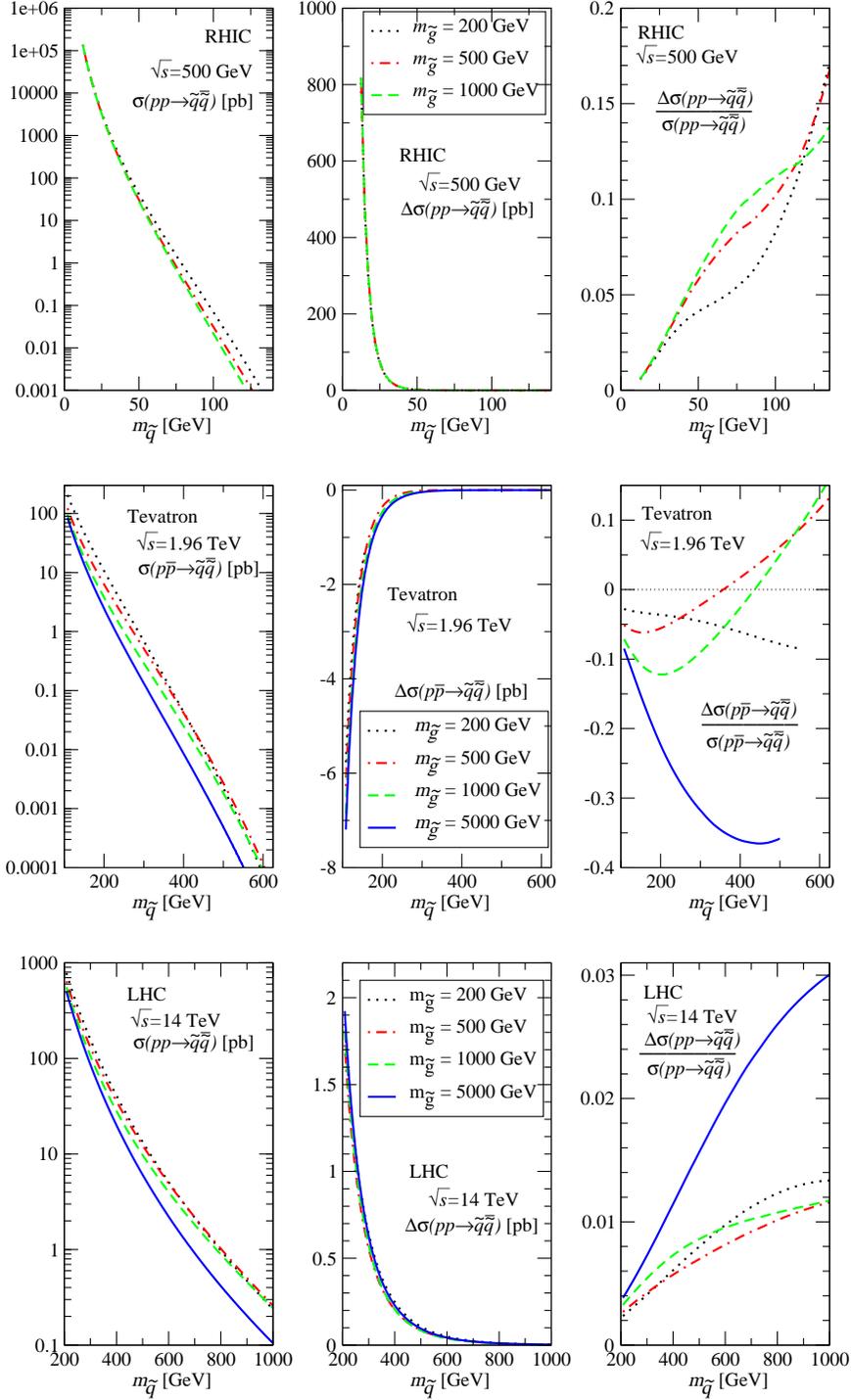

	\begin{center}
\epsfig{file=m_rhic_QQbar_sum_diff_asym.eps,width=11.5cm}
		\\ \vspace{0.4cm}
\epsfig{file=m_Teva_QQbar_sum_diff_asym.eps,width=11.5cm}
		\\ \vspace{0.4cm}
\epsfig{file=m_lhc_QQbar_sum_diff_asym.eps,width=11.5cm}
	\end{center}\caption{Total cross section, polarisation difference and asymmetry for the production of squark-antisquark for the three colliders LHC, Tevatron and RHIC.}
\label{m_lhc_sum_diff_asym_QQbar}
\end{figure}

\subsection{Sparticle Mass Dependence}

\begin{figure}[t]
	\begin{center}
\epsfig{file=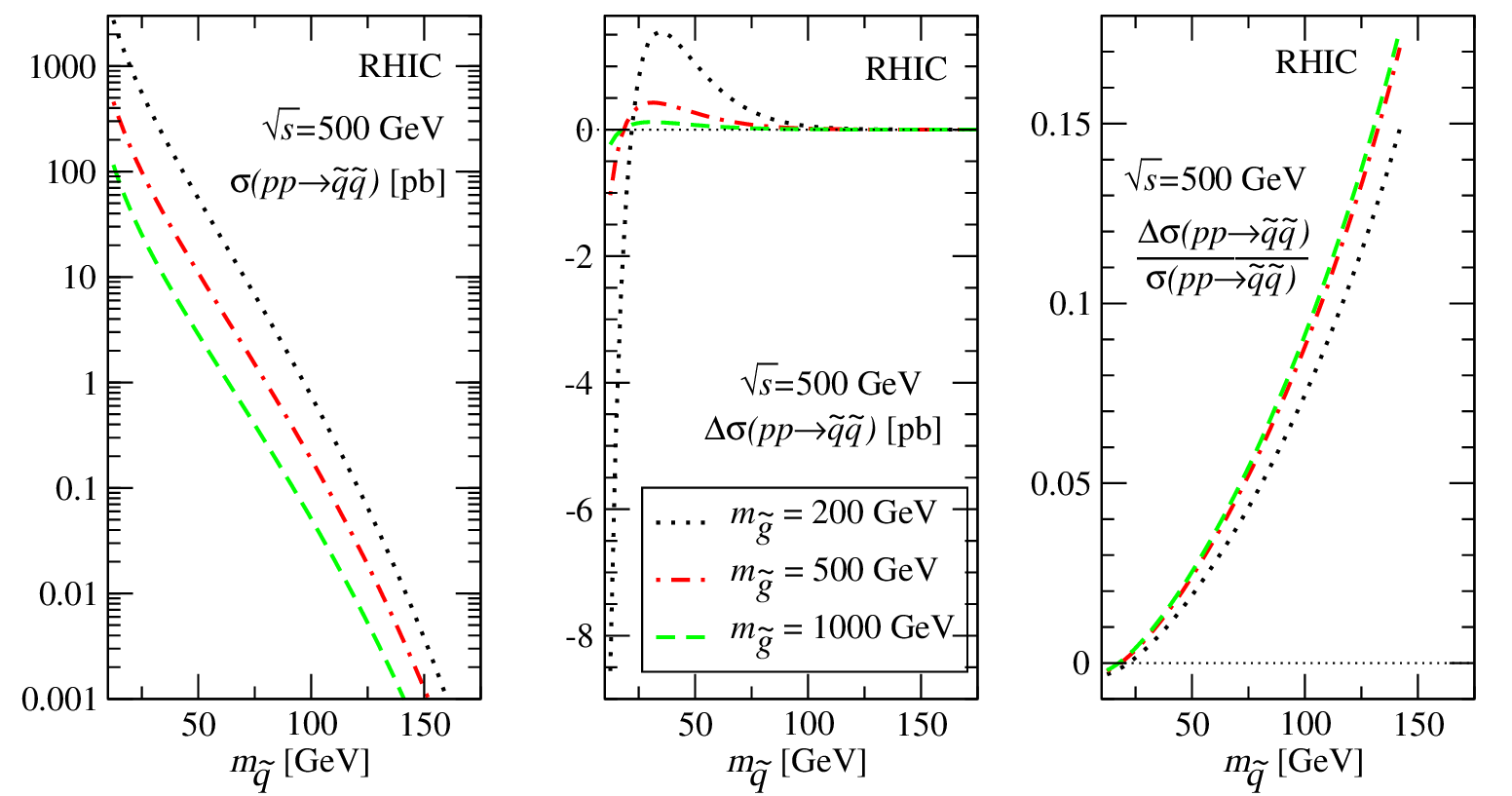,width=11.5cm}
		\\ \vspace{0.4cm}
\epsfig{file=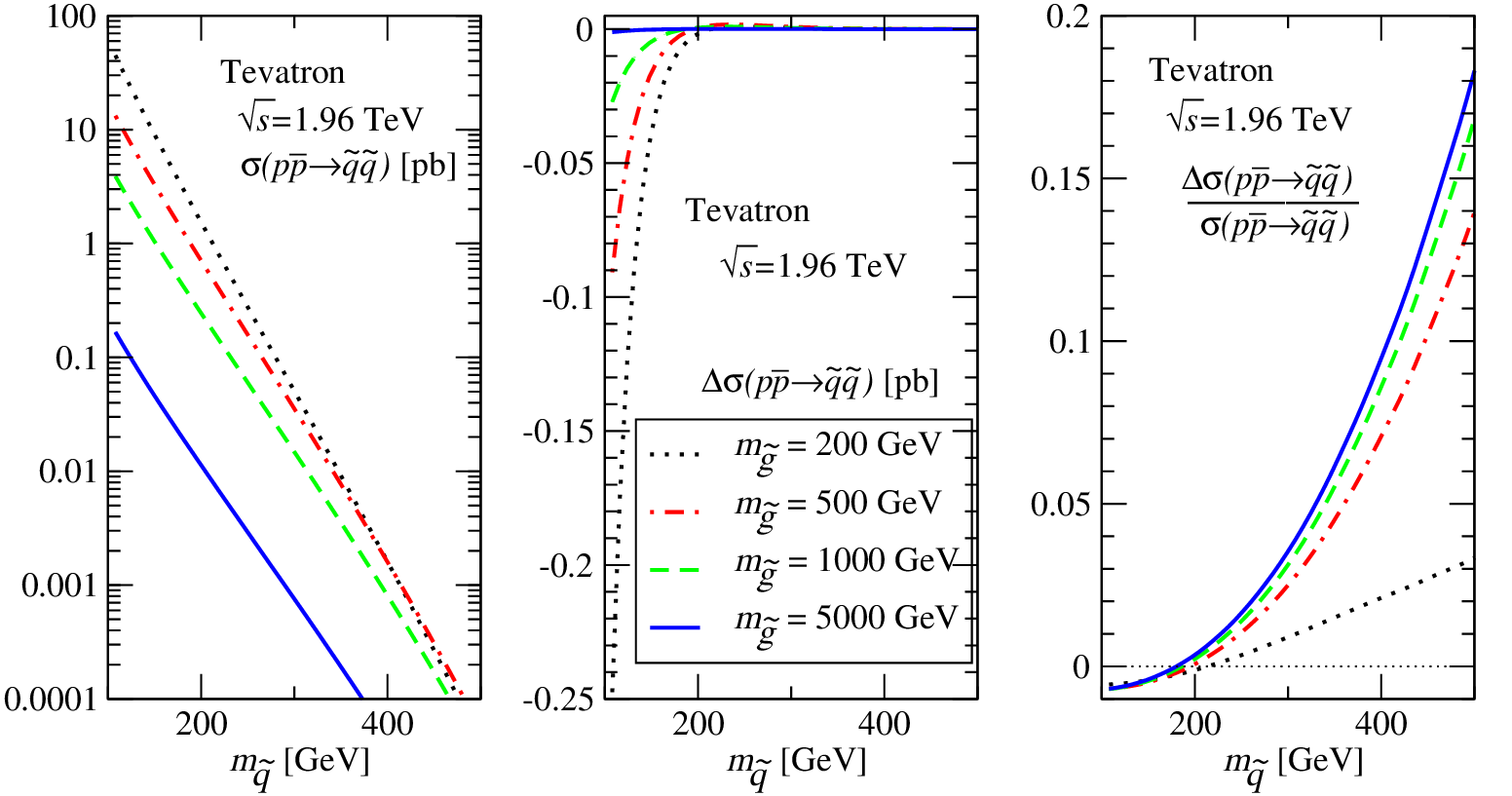,width=11.5cm}
		\\ \vspace{0.4cm}
\epsfig{file=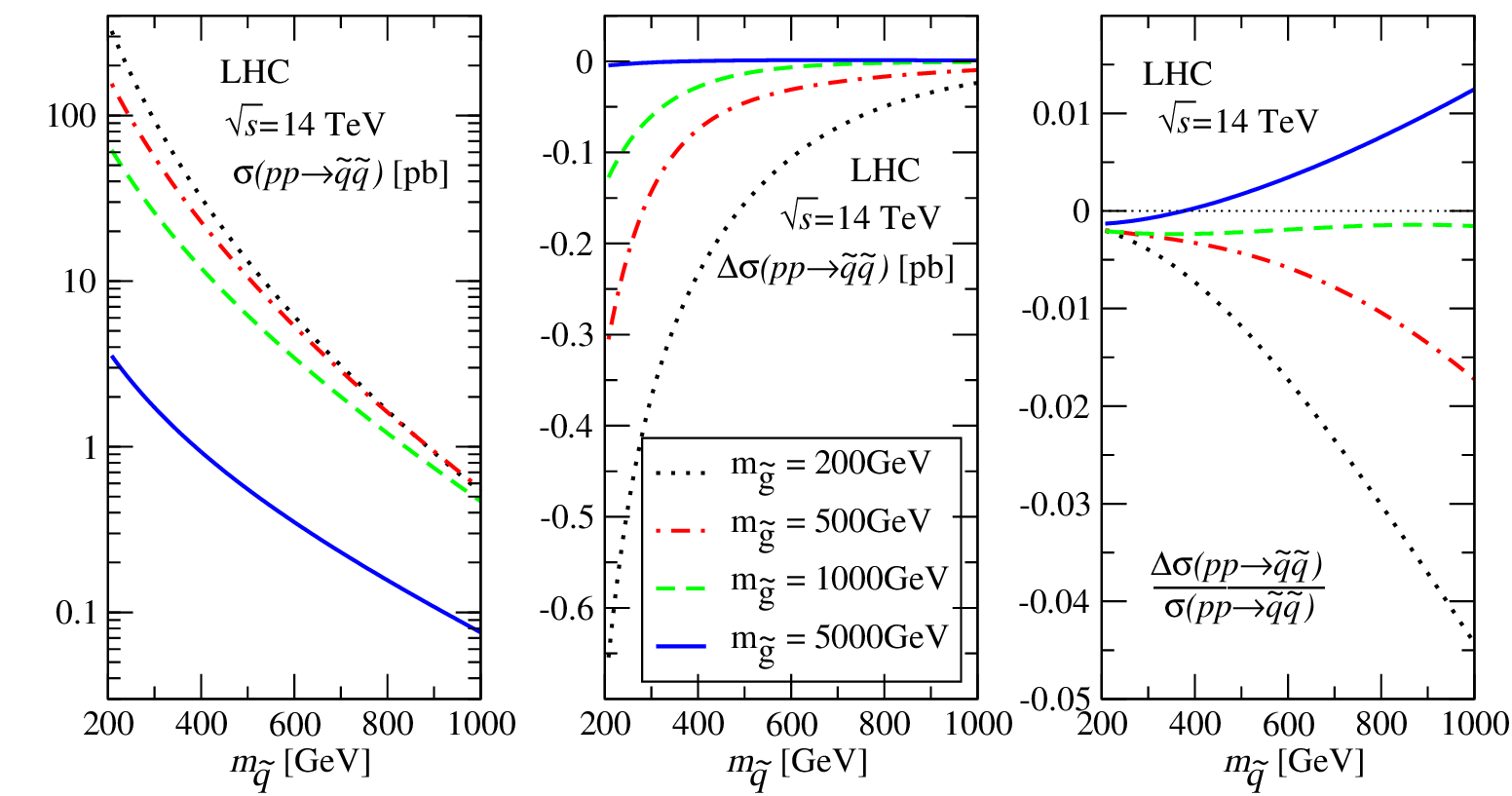,width=11.5cm}
	\end{center}
	\caption{Total cross section, polarisation difference and asymmetry for the production of squark-squark and antisquark-antisquark for the three colliders LHC, Tevatron and RHIC.}
\label{m_lhc_sum_diff_asym_QQ}
\end{figure}
\begin{figure}[t]
	\begin{center}
\epsfig{file=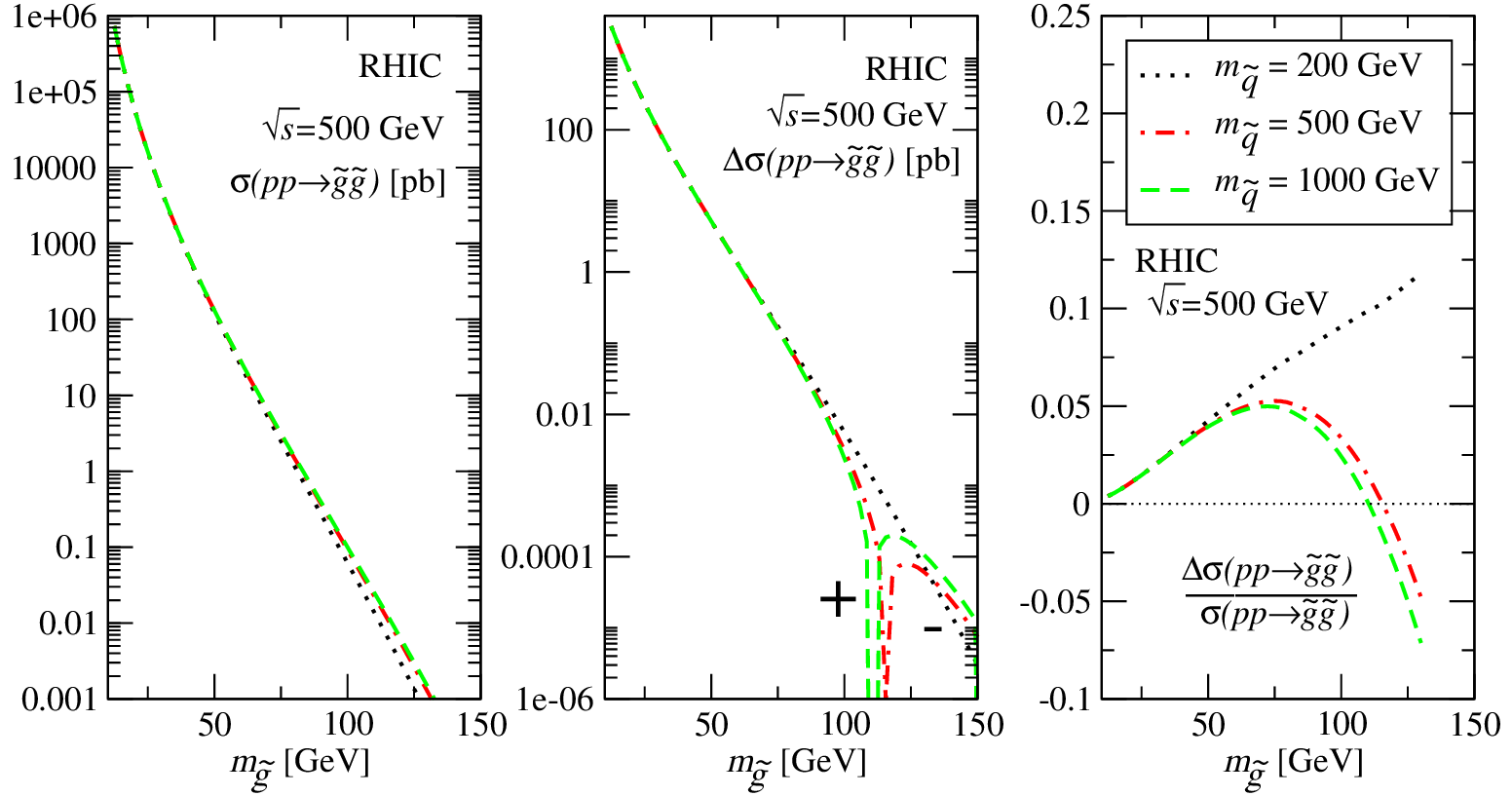,width=11.5cm}
		\\ \vspace{0.4cm}
\epsfig{file=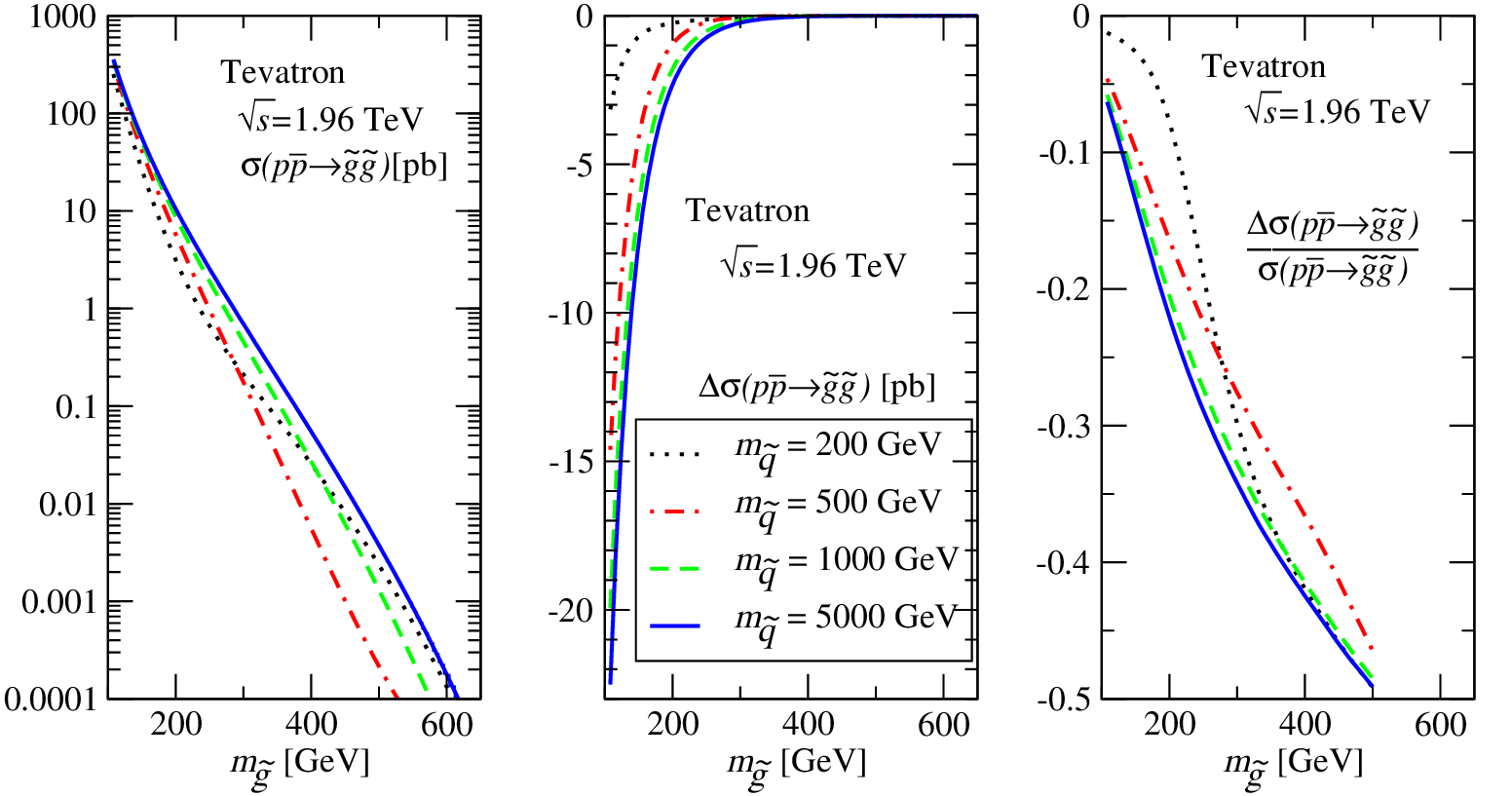,width=11.5cm}
		\\ \vspace{0.4cm}
\epsfig{file=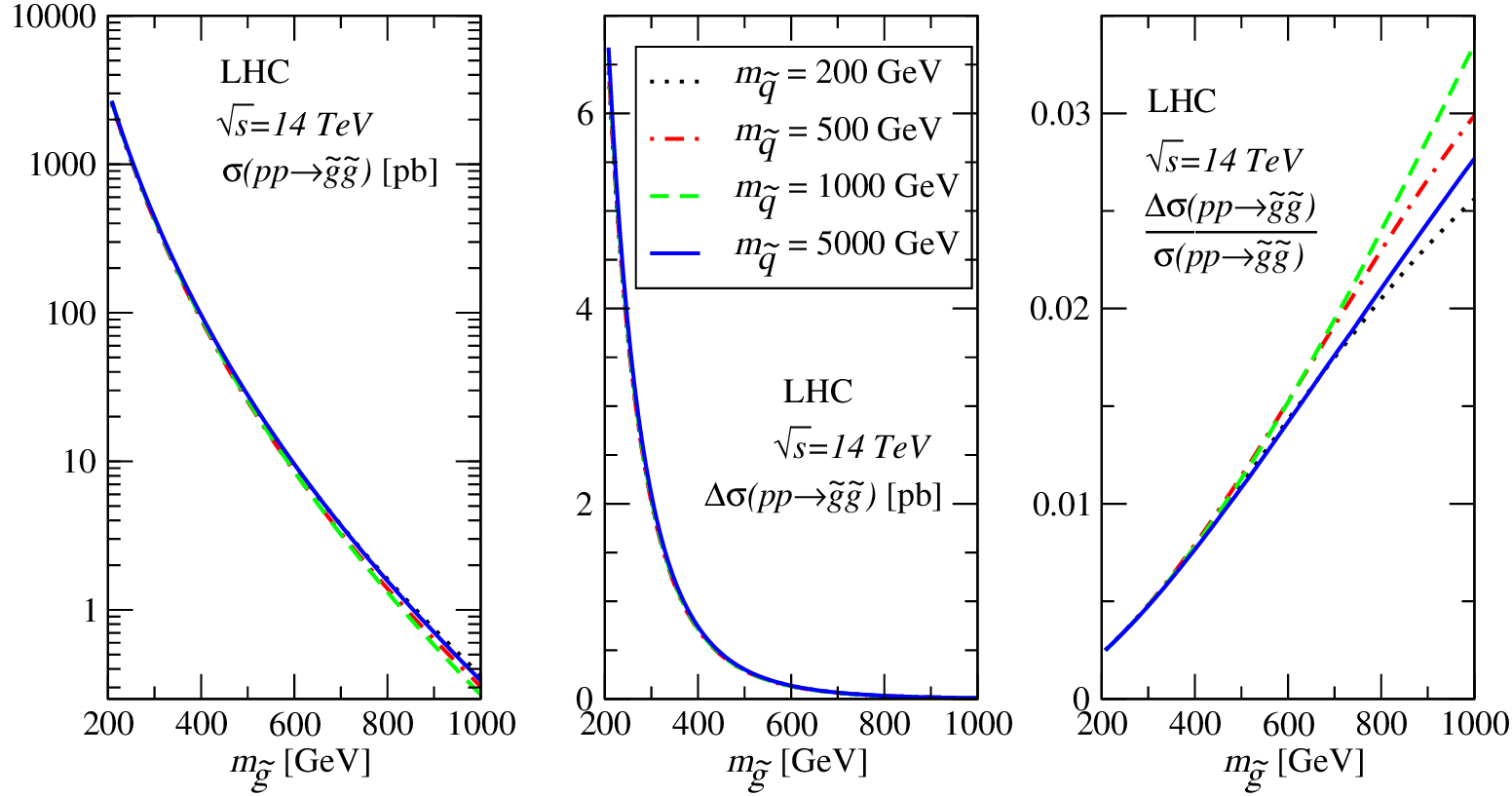,width=11.5cm}
	\end{center}
	\caption{Total cross section, polarisation difference and asymmetry for the production of gluino-gluino for the three colliders LHC, Tevatron and RHIC.}
\label{m_lhc_sum_diff_asym_GG}
\end{figure}
\begin{figure}[t]
	\begin{center}
\epsfig{file=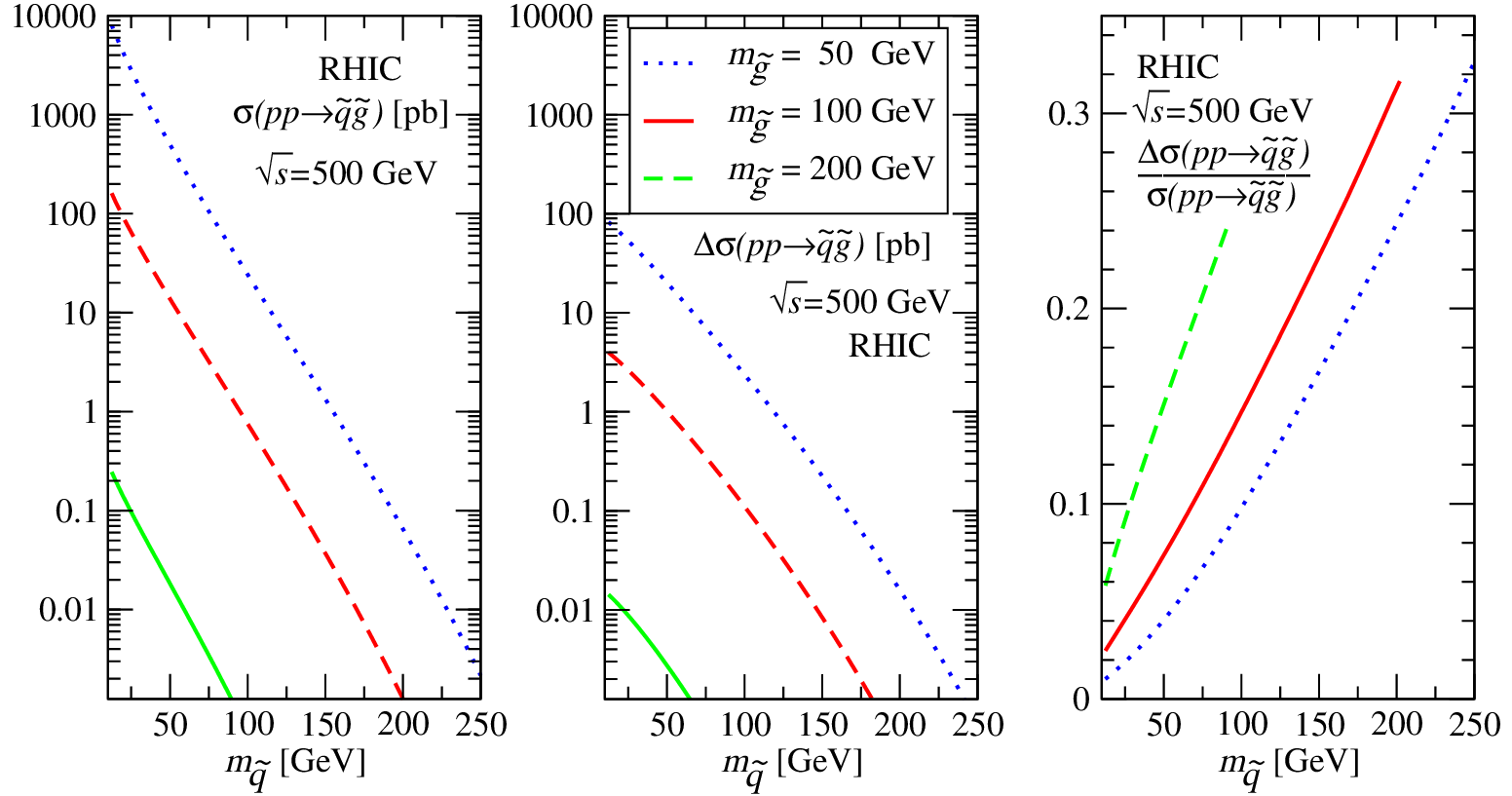,width=11.5cm}
		\\ \vspace{0.4cm}
\epsfig{file=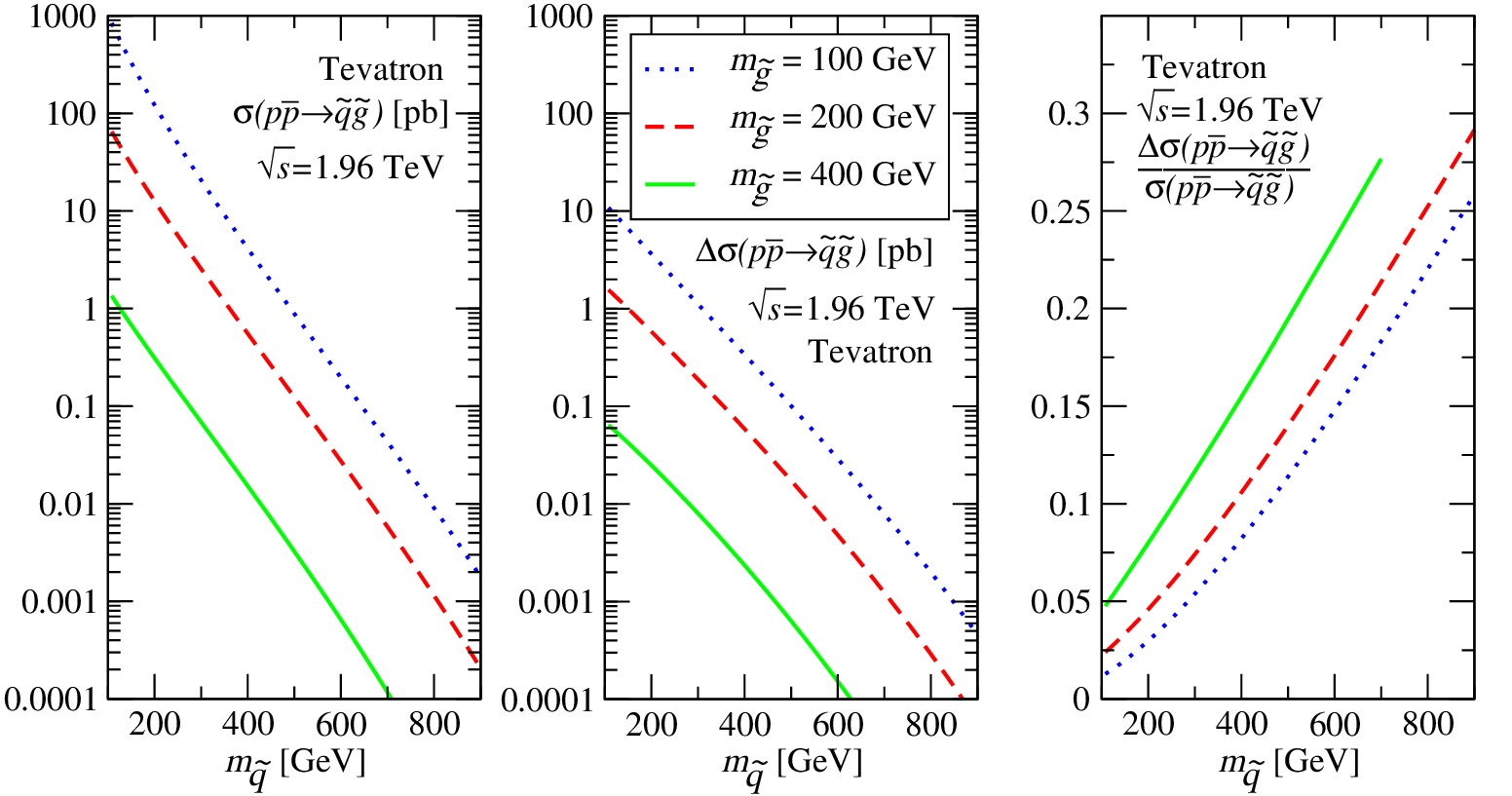,width=11.5cm}
		\\ \vspace{0.4cm}
\epsfig{file=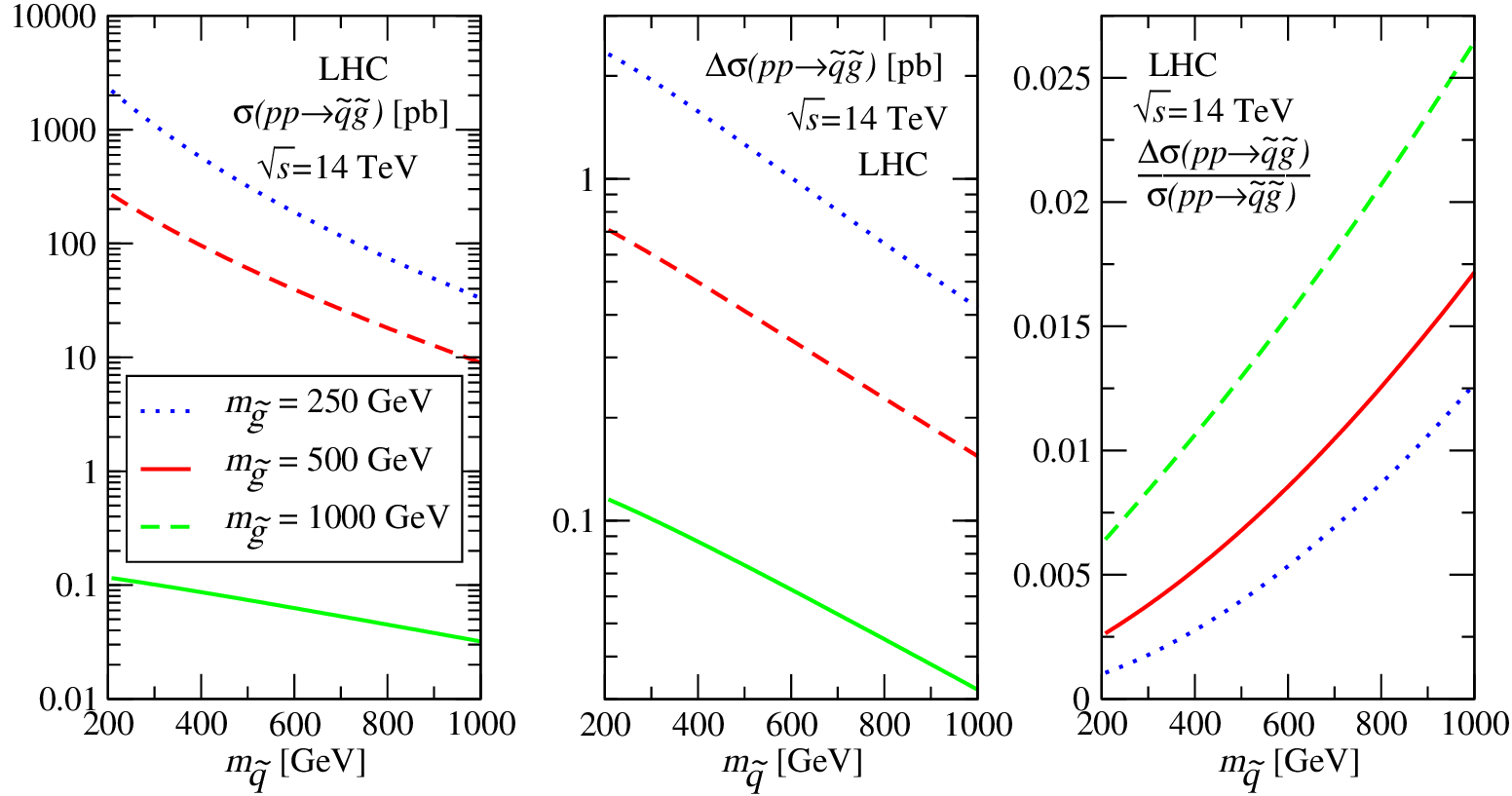,width=11.5cm}
	\end{center}
	\caption{Total cross section, polarisation difference and asymmetry for the production of squark(antisquark)-gluino for the three colliders LHC, Tevatron and RHIC.}
\label{m_lhc_sum_diff_asym_QG}
\end{figure}

The squark and gluino production cross sections and  
asymmetries depend on their masses. In the absence of 
observational evidence for supersymmetric particle production,
these masses are unknown. Within the MSSM, one can derive
the mass bounds~\cite{tevdata} $m_{\tilde{q}} > 250$~GeV, 
$m_{\tilde{g}}> 195$~GeV; in more complicated scenarios they are
significantly weaker. To exhibit the 
spin asymmetries one could expect, we scan the space of squark and 
gluino mass parameters. The results 
are displayed in 
Figures~\ref{m_lhc_sum_diff_asym_QQbar}--\ref{m_lhc_sum_diff_asym_QG}.

While the unpolarized and polarized cross sections fall rapidly 
with increasing sparticle masses, the asymmetry (ratio of 
polarized to unpolarized cross section) rises. This is because 
the ratio of polarized to unpolarized parton distributions is largest
for $x \simeq 1$. In all squark-antisquark and squark-squark
production asymmetries, one observes a sensitivity to the 
gluino mass, although the gluino appears in these processes only as 
an exchange particle. This feature could eventually be used to 
put constraints on the gluino mass from asymmetries in 
squark production, even if direct observation of the gluino is 
beyond the reach of the available collider. A similar determination from 
the unpolarized cross section would require a very precise absolute 
measurement of this cross section, and an equally precise theoretical 
understanding. 
In contrast, gluino production is only weakly dependent 
on the squark mass. 

We see that production asymmetries can reach 10\%--20\% for 
sparticle production at RHIC and Tevatron and and up to 4\% at the LHC
for the ranges of parameters plotted here. The reason for this lies again in
the mass dependence of the asymmetry; For higher sparticle masses,
asymmetries at LHC are even larger. However, as will be shown in 
Section~\ref{sec:errors} below, not enough events can be observed at these 
masses to allow a measurement of an asymmetry.

\subsection{Statistical Errors}
\label{sec:errors}

From the unpolarized cross sections we can estimate the statistical errors 
for the measurement of the spin asymmetries. We will use
the luminosity specifications postulated above for 
RHIC and the polarized options of Tevatron and LHC.
\begin{figure}[t]
	\begin{center}
		\epsfig{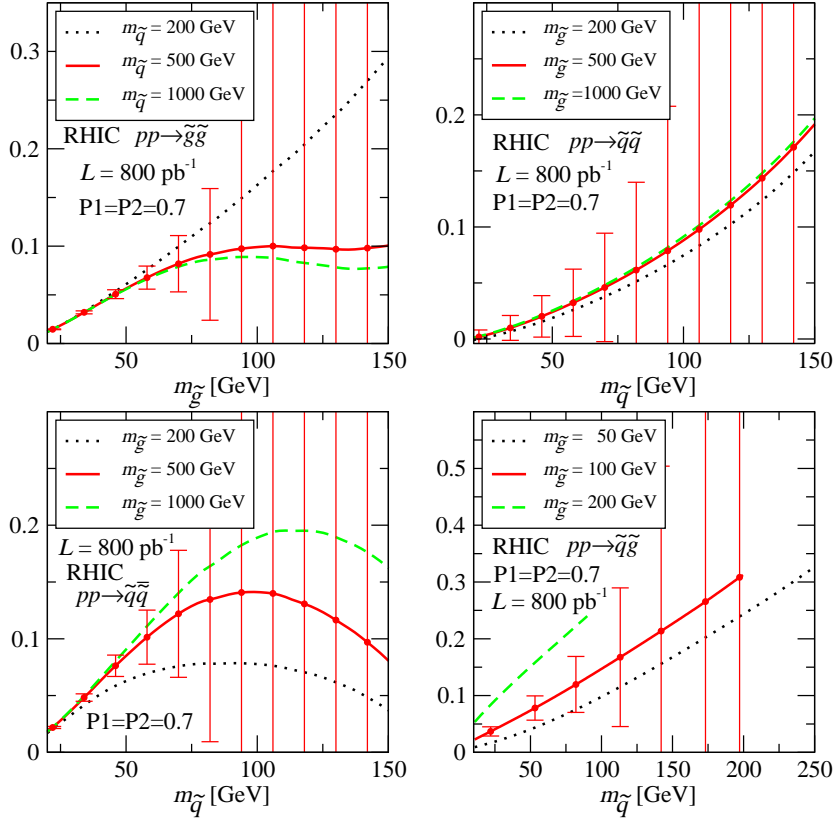}
	\end{center}
	\caption{Statistical error on the asymmetry for the RHIC. We have assumed a polarisation of 70\% and a total integrated luminosity of 800 pb$^{-1}$.}
\label{m_stat_rhic}
\end{figure}

\begin{figure}[t]
	\begin{center}
		\epsfig{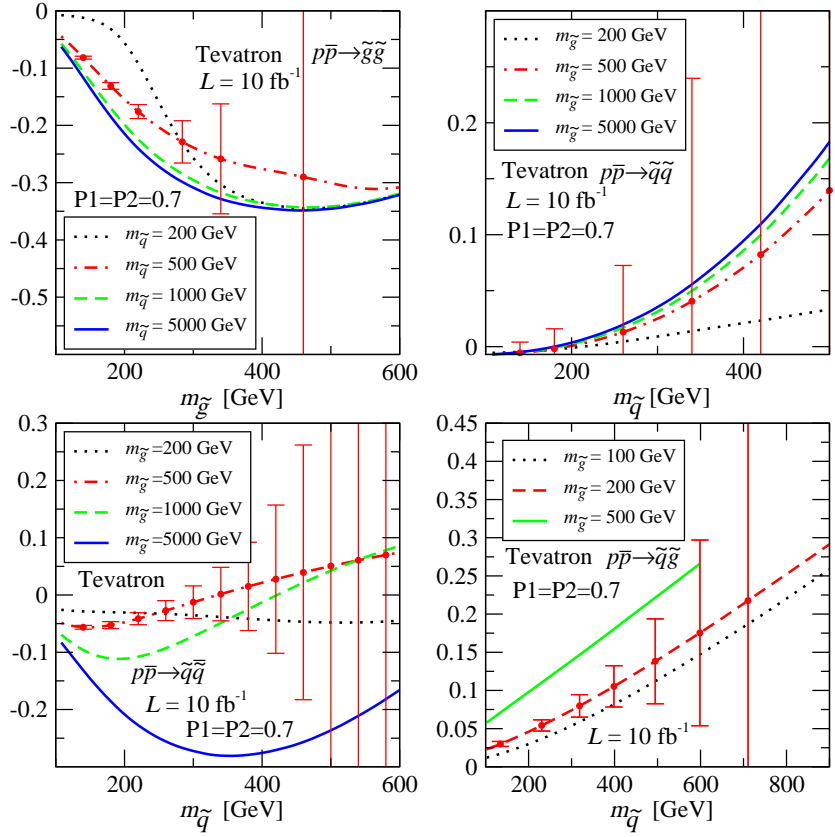}
	\end{center}
	\caption{Statistical error on the asymmetry for the Tevatron. We have assumed a polarisation of 70\% and a total integrated luminosity of 
10 fb$^{-1}$.}
\label{m_stat_Teva}
\end{figure}

\begin{figure}[t]
	\begin{center}
		\epsfig{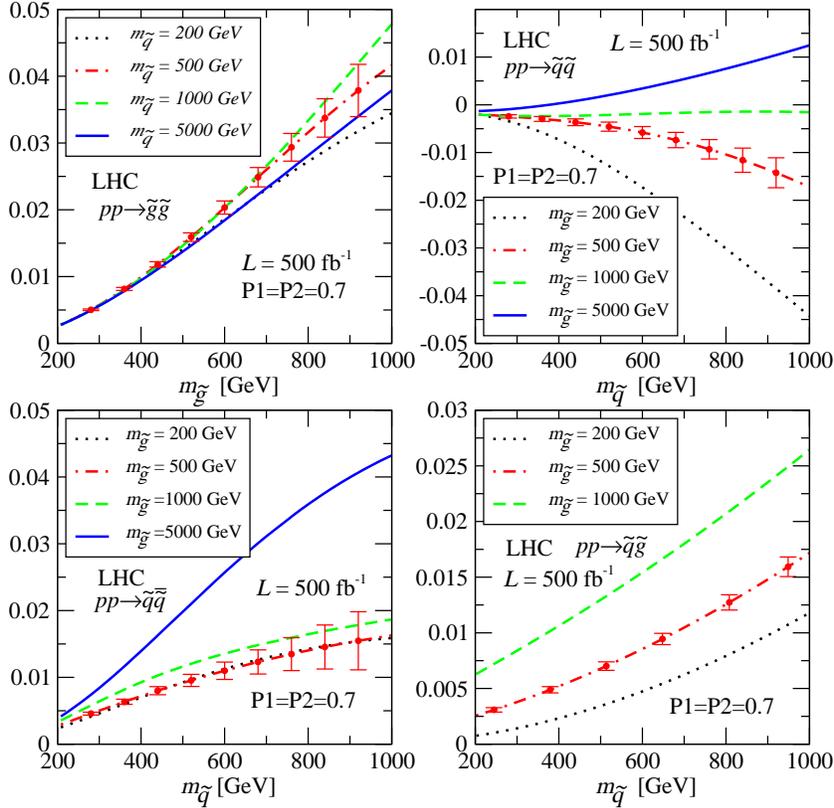}
	\end{center}
	\caption{Statistical error on the asymmetry for the LHC. We have assumed a polarisation of 70\% and a total integrated luminosity of 500 fb$^{-1}$.}
\label{m_stat_lhc}
\end{figure}

According to \cite{rhic}, the experimentally measured asymmetry is given by:
\begin{equation}
A=\frac{1 }{P_1P_2 }\frac{N_+ -N_- }{N_+ +N_- },
\end{equation}
where $P_1$ and $P_2$ are the polarisation of the two colliding beams 
and $N_{\pm}$ are defined as follows
\begin{equation}
N_+=N_{++} +N_{--} \qquad,\qquad N_-={N_{+-} +N_{-+} }.\nonumber
\end{equation}
$N_{rs}$ is the number of events counted with beams $s$- and $r$-polarised, 
normalised by the luminosity for this polarisation configuration. 
The statistical error $\Delta A$ for this measurement is given by:
\begin{equation}
\Delta A^2=\left< (A-<A>)^2\right>=<\delta A^2>.
\end{equation}
We assume that the results of the counts for $N_+$ and $N_-$ are 
Poisson distributed and that the error on $N_{\pm}$ due to the errors on the 
polarisation of the beams and on the luminosity are small compared to 
the deviation due to the Poisson statistic. The error on $N_{\pm}$ is then
\begin{displaymath}
\Delta N_{\pm}^2=N_{\pm},
\end{displaymath}

and for a small deviation we get for $\delta A$:
\begin{eqnarray}
\delta A^2&=&\frac{4 }{P_1^2P_2^2 }\frac{\Delta N_+^2 N_-^2 +\Delta N_-^2 N_+^2-2\Delta N_-\Delta N_+N_+N_- }{(N_+ +N_-)^2} .\nonumber
\end{eqnarray}
The expectation value of this expression gives $\Delta A^2$:
\begin{equation}
\Delta A^2=<\delta A^2>=\frac{4 }{P_1^2P_2^2 }\frac{<\Delta N_+>^2 N_-^2 +<\Delta N_-^2> N_+^2}{(N_+ +N_-)^2}.
\end{equation}
The final result for the statistical error of the asymmetry is obtained
by inserting the deviation $\Delta N_{+/-}$ and expressing the result as a function of the asymmetry $A$ and $N=N_++N_-$:
\begin{equation}\label{stat_error}
\Delta A^2=\frac{4(N_+N_-)}{N^2P_1^2P_2^2}=\frac{1}{NP_1^2P_2^2}-\frac{A^2}{N^2}.
\end{equation}
The expected statistical errors are collected 
in Figures \ref{m_stat_rhic}-\ref{m_stat_lhc}. 

For all three collider scenarios, the statistical errors grow 
with increasing sparticle masses such that statistically 
significant measurements are especially possible for low masses.

At RHIC the situation is quite favourable for $m_{\G/\Q}\lapprox 75$ GeV, except
in squark-squark production. This process is
a pure gluino $t$-channel exchange which is suppressed by the gluino mass.
%Consequently, the asymmetry in this channel is not measurable. 

At Tevatron, with a total integrated luminosity of 10 fb$^{-1}$, 
the statistical error starts getting important in the 
region $m_{\G/\Q}\approx 350$ GeV; the exception is 
squark-squark production where the statistical error is larger 
than the asymmetry everywhere. 
At Tevatron, squark-squark production is particularly suppressed
because the quark-quark luminosity is lower at a 
proton-antiproton collider than at a proton-proton collider. 

At LHC, 
with a total integrated luminosity of 500 fb$^{-1}$, 
all channels (with the exception of 
squark-squark production with 
large gluino mass) display only moderate errors even up to 
sparticle masses of 1000~GeV. Therefore a measurement of 
spin asymmetries at LHC would be especially interesting, despite their
apparent smallness.

In a realistic measurement it will be very difficult to distinguish between
squark-antisquark and squark-squark final states and both 
might have to be summed over. The summed cross sections are mostly dominated by 
squark-antisquark final states, and the 
combined asymmetry would be of similar magnitude as
 the squark-antisquark asymmetry~\cite{dipl}.

\section{Transverse Momentum and Rapidity Spectra}
\label{sec:pt}

In the previous section, we 
have established measurable spin asymmetries in inclusive 
squark and gluino production for a wide range of sparticle masses.
In a  measurement, one is however often not able to 
detect particles over the fully inclusive final state phase space
but only in restricted ranges in the 
final state transverse momentum and rapidity.
It is therefore important to study the dependence of the 
cross sections and asymmetries on the sparticle rapidity and transverse 
momentum. 

\begin{figure}[t]
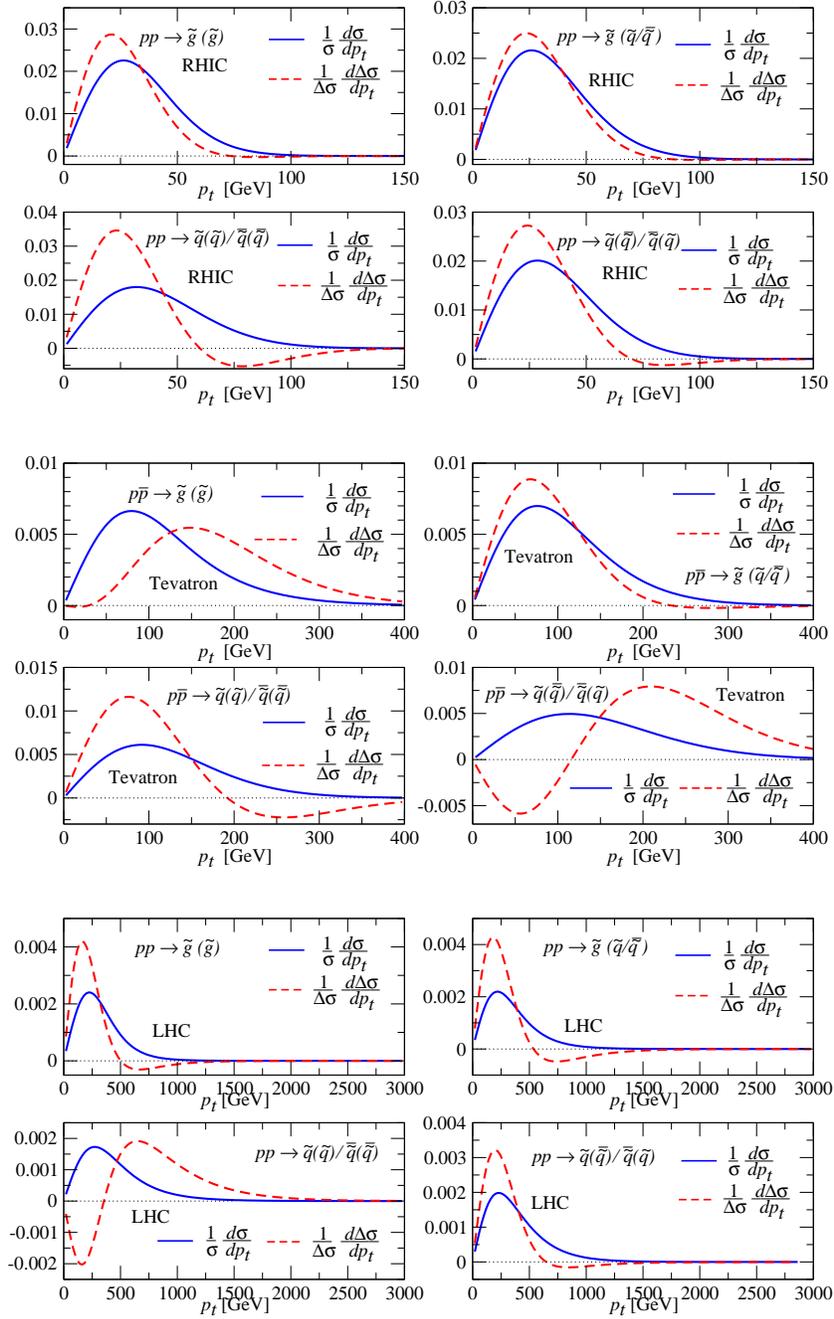

	\begin{center}
		\epsfig{file=trans_rhic_sum_diff.eps,width=11cm}
		\\
		\vspace{0.5cm}
		\epsfig{file=trans_Teva_sum_diff.eps,width=11cm}
		\\
		\vspace{0.5cm}
		\epsfig{file=trans_lhc_sum_diff.eps,width=11cm}
	\end{center}
	\caption{Normalised transverse momentum distribution for the three colliders LHC, Tevatron and RHIC. The mass parameters are for the LHC ($\sqrt{S}=14$ TeV) $m_{\G}=500$ GeV, $m_{\Q}=600$ GeV, for the Tevatron ($\sqrt{S}=1.96$ TeV) $m_{\G}=200$ GeV, $m_{\Q}=280$ GeV and for the RHIC ($\sqrt{S}=500$ GeV) $m_{\G}=80$ GeV, $m_{\Q}=100$ GeV. }
\label{trans_all_sum_diff}
\end{figure}
\begin{figure}[t]
	\begin{center}
\epsfig{file=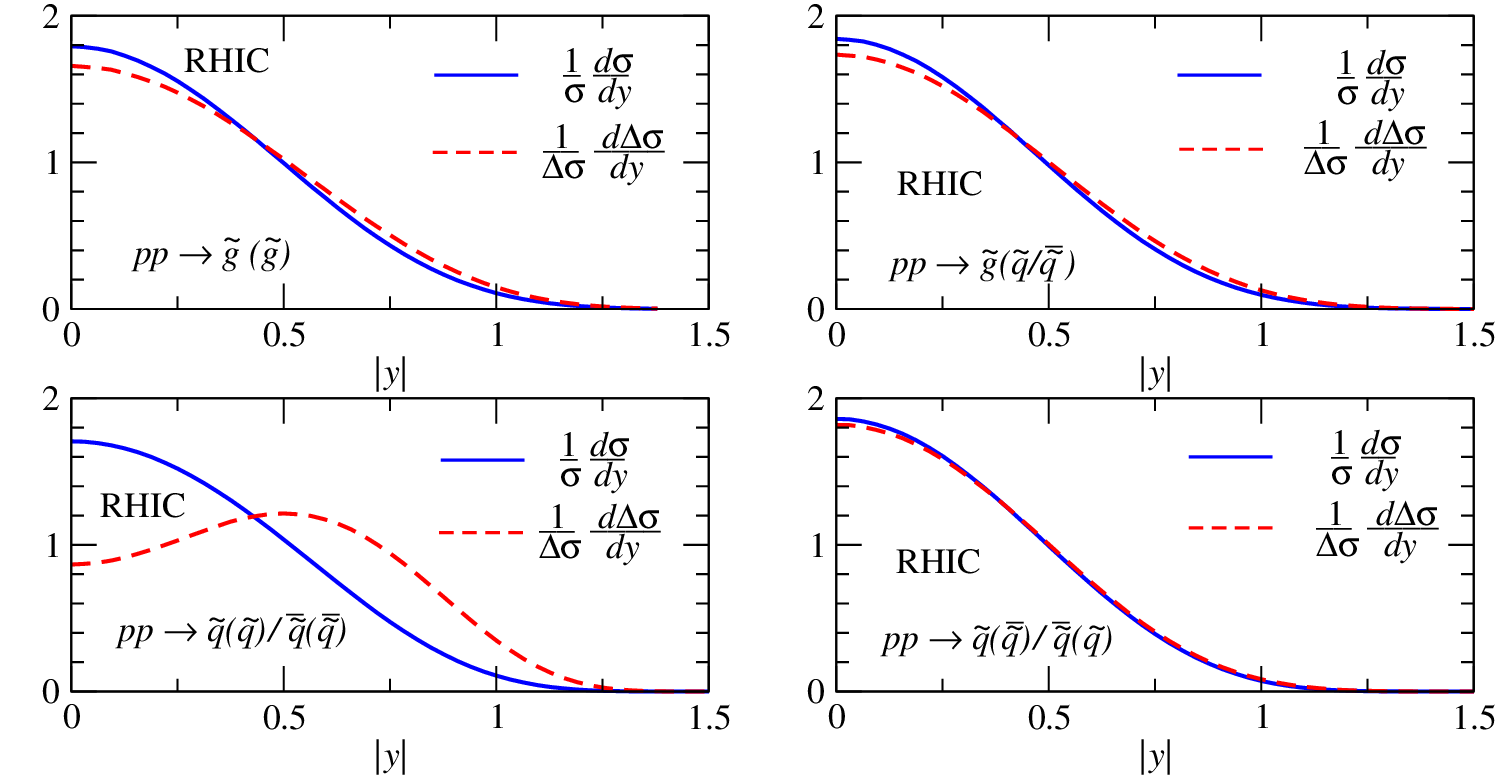,width=11cm}
		\\
		\vspace{0.5cm}
\epsfig{file=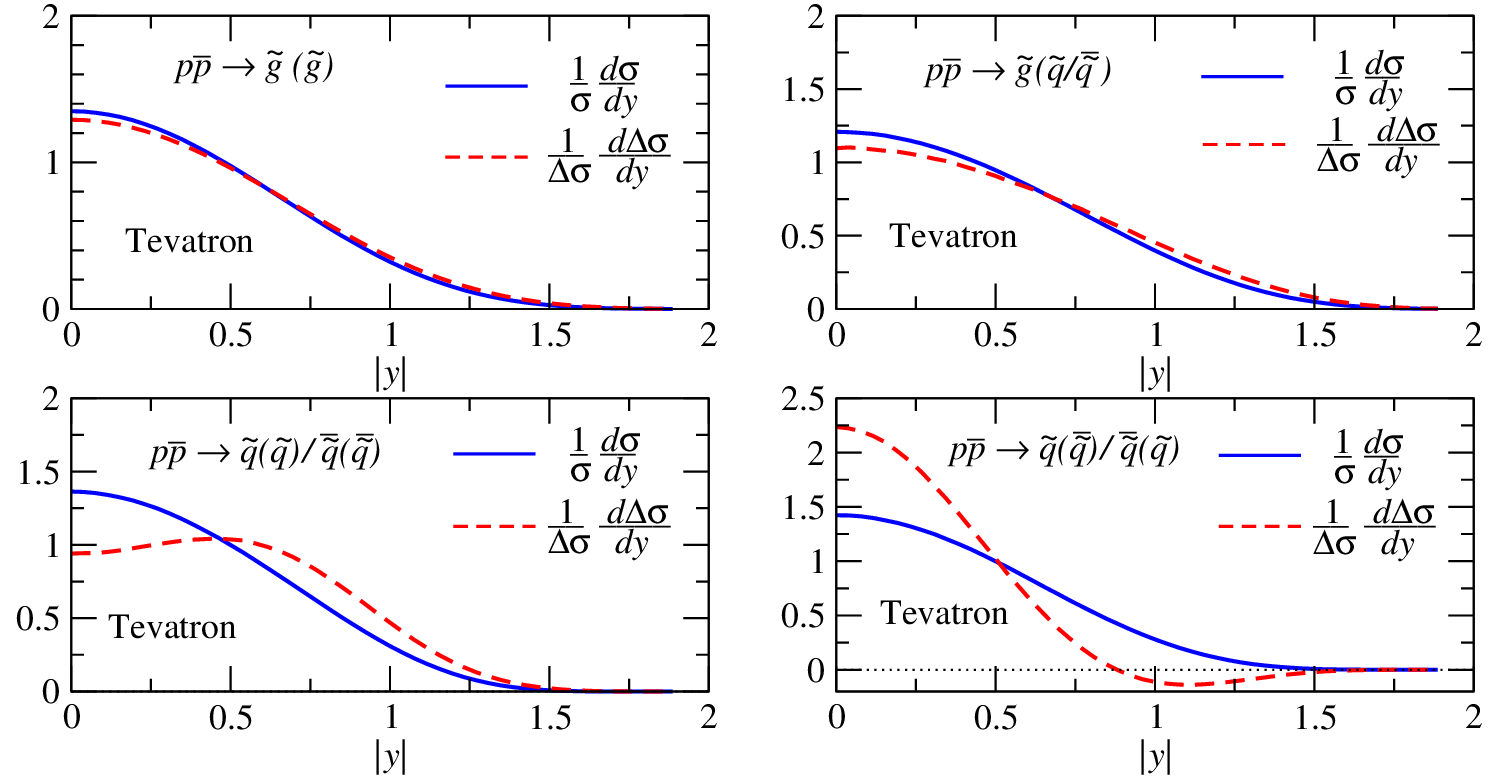,width=11cm}
		\\
		\vspace{0.5cm}
\epsfig{file=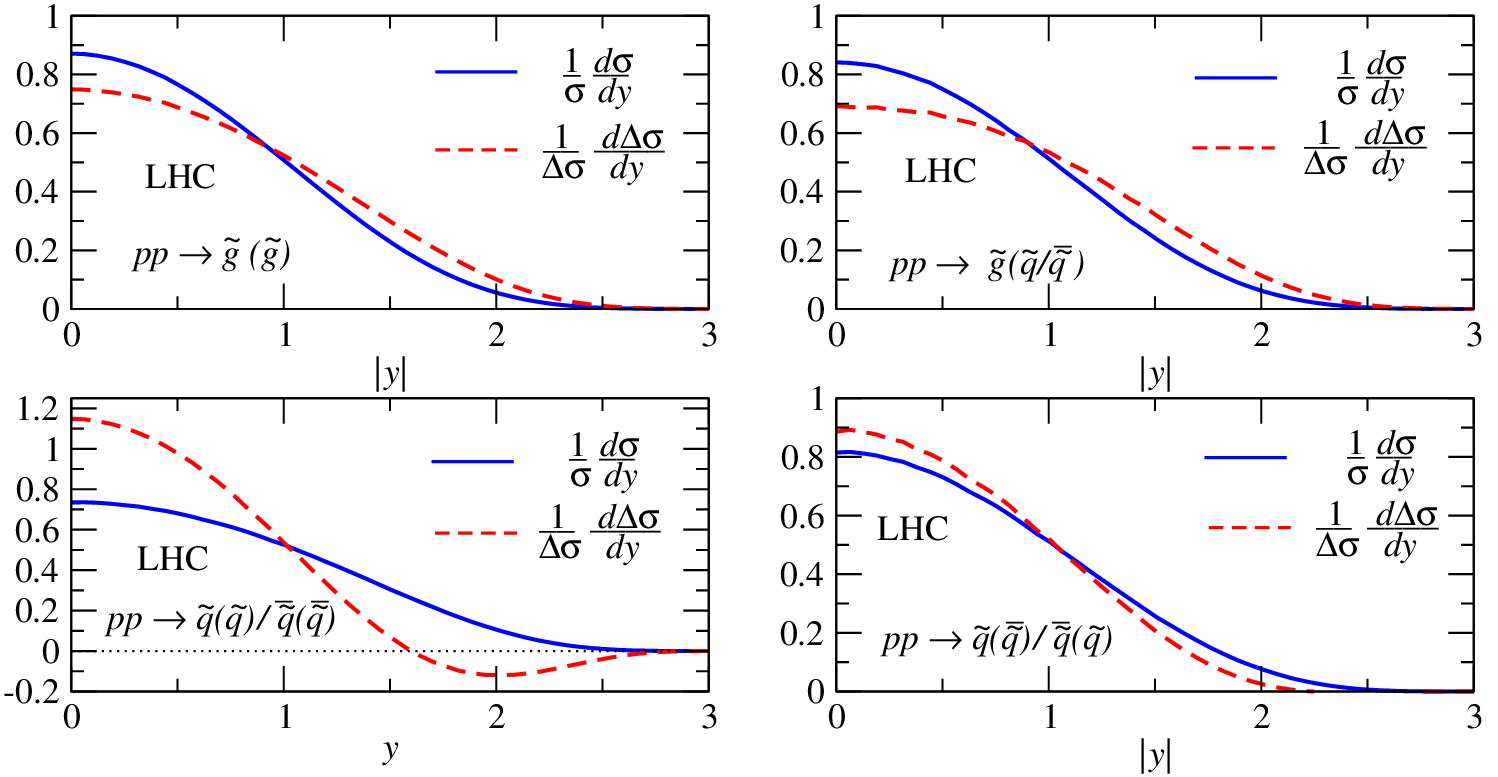,width=11cm}
 	\end{center}
	\caption{Normalised rapidity distributions for the three colliders LHC, Tevatron and RHIC. The mass parameters are for the LHC ($\sqrt{S}=14$ TeV) $m_{\G}=500$ GeV, $m_{\Q}=600$ GeV, for the Tevatron ($\sqrt{S}=1.96$ TeV) $m_{\G}=200$ GeV, $m_{\Q}=280$ GeV and for the RHIC ($\sqrt{S}=500$ GeV) $m_{\G}=80$ GeV, $m_{\Q}=100$ GeV. }
\label{rap_all_sum_diff}
\end{figure}
The transverse momentum distribution of unpolarized and polarized 
cross sections are displayed in Figure~\ref{trans_all_sum_diff}. 
We observe that the 
majority of the sparticles are produced with sizable transverse momentum of 
about $(0.2\ldots 0.5)m_{\Q / \G}$. Moreover, for almost all final states, we 
observe that the transverse momentum distribution of unpolarized and polarized 
cross sections differ substantially. At the LHC, all polarized distributions 
change their sign between small and large transverse momenta. A similar 
feature is observed in some distributions at RHIC and the Tevatron. This 
behaviour reflects in part the structure of the polarized matrix elements, 
but also the form of the polarized parton distribution functions. 
\begin{figure}[t]
	\begin{center}
		\epsfig{file=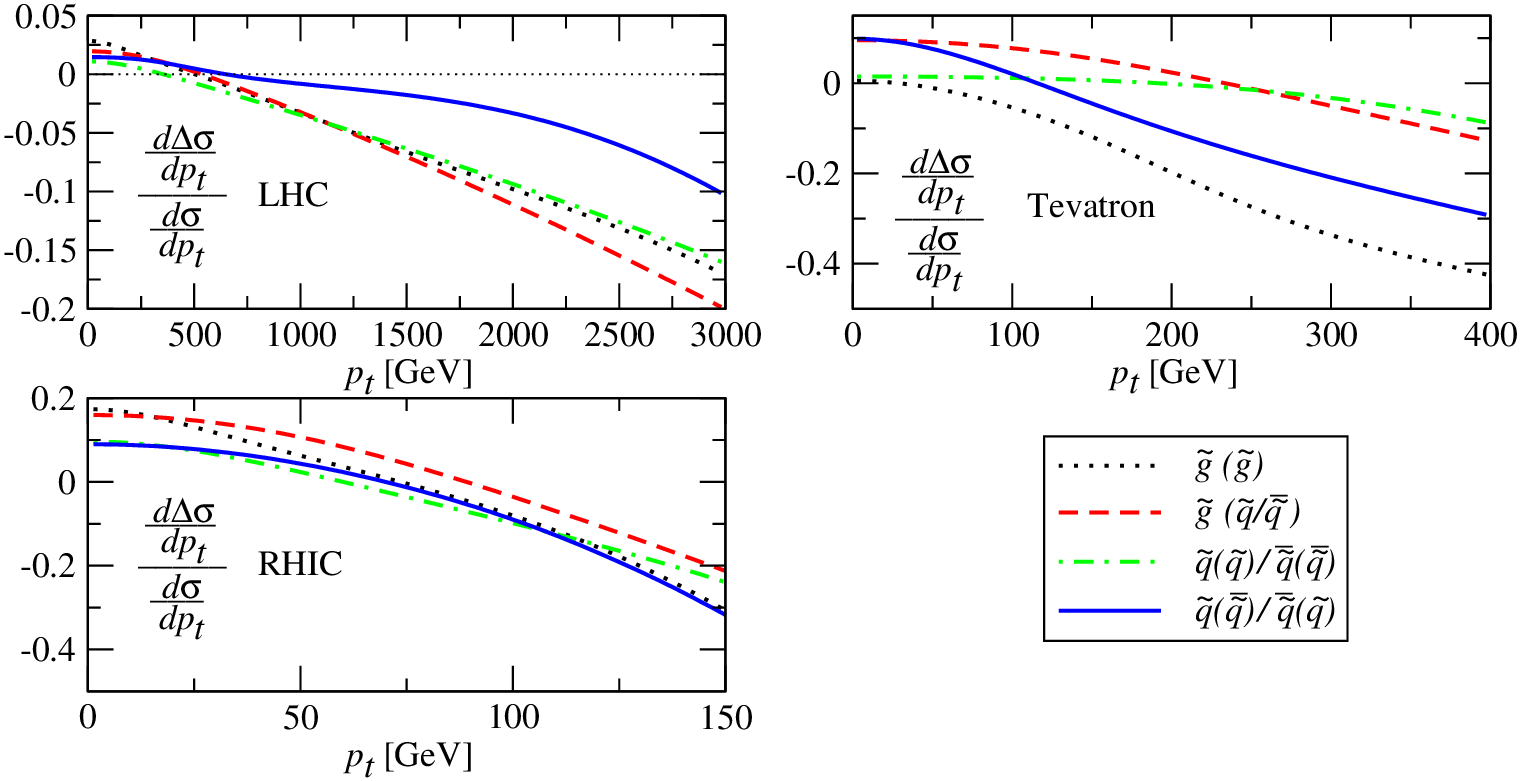,width=12cm}
	\end{center}
	\caption{Asymmetry in the transverse momentum distribution for the three colliders LHC, Tevatron and RHIC. The mass parameters are for the LHC ($\sqrt{S}=14$ TeV) $m_{\G}=500$ GeV, $m_{\Q}=600$ GeV, for the Tevatron ($\sqrt{S}=1.96$ TeV) $m_{\G}=200$ GeV, $m_{\Q}=280$ GeV and for the RHIC ($\sqrt{S}=500$ GeV) $m_{\G}=80$ GeV, $m_{\Q}=100$ GeV. }
\label{trans_asym}
	\begin{center}
		\epsfig{file=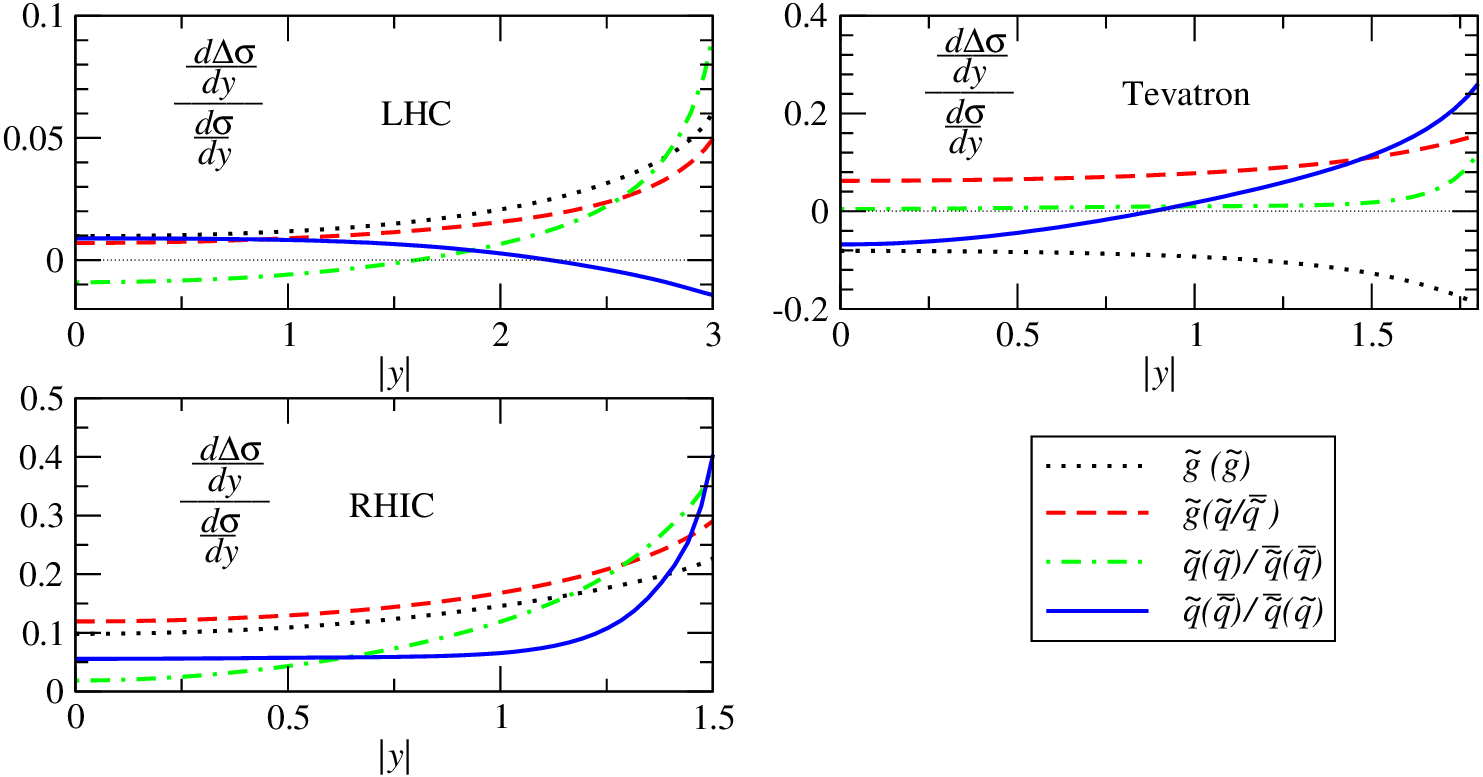,width=12cm}
	\end{center}
	\caption{Asymmetry in the rapidity distribution for the three colliders LHC, Tevatron and RHIC. The mass parameters are for the LHC ($\sqrt{S}=14$ TeV) $m_{\G}=500$ GeV, $m_{\Q}=600$ GeV, for the Tevatron ($\sqrt{S}=1.96$ TeV) $m_{\G}=200$ GeV, $m_{\Q}=280$ GeV and for the RHIC ($\sqrt{S}=500$ GeV) $m_{\G}=80$ GeV, $m_{\Q}=100$ GeV. }
\label{rap_asym}
\end{figure}

At large transverse momenta, 
most polarized distributions fall off slower than their 
unpolarized counterparts. 
The explanation for this feature lies in the ratio between polarized and 
unpolarized parton distributions which increases towards larger values of $x$,
corresponding to large values of the transverse momentum. 

The rapidity distributions, Figure~\ref{rap_all_sum_diff},
 are more uniform than the 
transverse momentum distributions. Most polarized distributions follow their 
unpolarized counterparts rather closely, showing in general a 
slower decrease towards larger rapidities. 

The features of polarized and unpolarized 
transverse momentum and rapidity distributions become particularly visible in 
corresponding asymmetries, Figures~\ref{trans_asym} and~\ref{rap_asym}.
It can be seen that the transverse momentum spectrum of the spin asymmetry 
increases towards larger transverse momenta for all colliders and all final 
states. Several spectra display a zero. On the contrary, the rapidity
 dependence of the spin asymmetries is rather flat, especially in the 
experimentally most relevant central rapidity region.

\section{Conclusions and Outlook}
\label{sec:conc}

In this paper, we have studied spin asymmetries in squark and 
gluino production at polarized hadron colliders to leading order in 
QCD. We consider three 
scenarios, the currently operational polarized proton-proton collider 
RHIC and hypothetical polarized versions of the Tevatron and the LHC. 
For all three colliders, we find sizable spin asymmetries. Assuming the 
design luminosities and beam polarization of 70\%,  the asymmetries are
statistically measurable for sparticle masses up to 
75~GeV at RHIC, 350~GeV at the Tevatron and well above 1~TeV at LHC,
provided experimental systematic uncertainties on them can be 
kept under control.

If supersymmetry is realized in nature, one would expect to see 
visible signatures of squarks and gluinos already 
in unpolarized hadronic collisions 
at the Tevatron or the LHC, depending on the particle masses. 
A subsequent
measurement of spin asymmetries at potential polarized versions of these 
colliders could be vital to 
give detailed insights into the new physics
and may distinguish between different supersymmetric models (or any other
new physics idea). For instance,
if (heavy) quarks and related squarks have similar masses and therefore similar
production rates, the corresponding asymmetries are different. An asymmetry
measurement could thus exclude a light bottom squark effectively 
\footnote{As mentioned previously, this possibility is presumably already
excluded.} or distinguish between a heavy fourth generation quark and a
supersymmetric particle. It would also be quite significant if the sparticle
masses are quite different from each other.
For instance, as shown in the text,
spin asymmetries in squark-antisquark production could be used 
to infer the gluino mass, even if direct gluino observation is beyond 
the kinematical reach of the collider. 
Apart from the additional
information it provides, the spin asymmetry can also be measured with
much less uncertainty than cross sections. As explained in the text,
spin asymmetries are insensitivity to absolute normalisation, which turns out 
to be a serious challenge at hadron colliders. We expect
to have detailed knowledge of the required
polarized parton distributions after measurements at 
RHIC and COMPASS have been completed.
Our results contribute to making a physics case for 
a new polarized hadron collider at energies above 
RHIC or for an upgrade of an existing and planned high energy colliders
to polarized hadrons. 

There are further phenomenological 
applications. One particular aspect of using polarized initial states is 
that they offer the possibility to obtain event samples enriched either 
in $L$- or $R$-squarks which can not be discriminated at unpolarized hadron 
colliders.

A precise study of spin asymmetries in squark and gluino
production would not only require an improved understanding of the 
spin-dependent parton distributions, but also the inclusion of 
next-to-leading order corrections to the partonic scattering 
processes. These are available for unpolarized squark and gluino 
production~\cite{been}. The calculation of the spin-dependent 
NLO corrections would resemble the NLO calculation of heavy quark 
production at polarized hadron colliders, which was completed some 
time ago~\cite{bs}. In fact, as pointed out already in~\cite{bs}, 
the NLO corrections to the polarized cross section for 
$gg\to \G\G$ could already be obtained from 
the results on spin-dependent heavy quark production.

One possible environment to study polarized partonic scattering processes 
would be photon-photon interactions at a future linear collider~\cite{tesla}, 
either realized by tagging onto bremsstrahlung photons or with a dedicated 
photon-photon programme using high energy photon beams generated by laser 
backscattering. In both options, the photon beams would be polarized. 
The polarized cross sections for squark and gluino production in 
direct and resolved photon interactions can be readily obtained from 
the formulae presented here. The polarized parton distributions in the
photon are at present completely unconstrained by experimental data, 
theoretical models for them exist and put upper and lower boundaries on 
their magnitude~\cite{svgam}. Knowledge on them could furthermore be improved 
by studying spin asymmetries in standard model reactions in photon-photon 
collisions.

\section*{Acknowledgements}
We would like to thank Philip Ratcliffe for pointing our attention 
to~\cite{ratcliffe} and for useful discussions.  
This work was supported by the Swiss National Funds 
(SNF) under contract 200021-101874.

\end{document}